\documentclass[10pt,a4paper]{article}
\usepackage{amsfonts}
\usepackage{verbatim}
\usepackage{color}
\usepackage{epsfig}
\usepackage{amsfonts}
\usepackage{amssymb}
\usepackage{amsmath}
\usepackage{amscd}
\usepackage{amsthm}
\usepackage[T1]{fontenc}
\usepackage[utf8]{inputenc}
\usepackage{graphicx}
\usepackage[english]{babel}
\newcommand{\mathsym}[1]{{}}

\definecolor{azul}{rgb}{0.1,0.2,1} 
\definecolor{verde}{rgb}{0.4,0.6,0.4}
\definecolor{bordo}{rgb}{0.8,0.5,0.3}
\newtheorem*{1 Kir}{1$^{\circ}$ Ley de Kirchhoff}
\newtheorem*{2 Kir}{2$^{\circ}$ Ley de Kirchhoff}

\begin{document}

\title{Equivalence between classical and quantum dynamics. Neutral kaons and electric circuits}
\author{M. Caruso, H. Fanchiotti, C.A. Garcia Canal
\vspace*{1cm}\\Departamento de Física, Facultad de Ciencias Exactas, IFLP,
\\Universidad Nacional de La Plata, C.C.67, La Plata (1900), Argentina }
\maketitle

\begin{abstract}
\par
An equivalence between the $\mathrm{Schr\ddot{o}dinger}$ dynamics of a quantum system with a finite number of basis states  and a classical dynamics is presented. The equivalence is an isomorphism that connects in univocal way both dynamical systems. We treat the particular case of neutral kaons and found a class of electric networks uniquely related to the kaon system finding the complete map between the matrix elements of the effective Hamiltonian of kaons and those elements of the classical dynamics of the networks. As a consequence, the relevant $\epsilon$ parameter that measures CP violation in the kaon system is completely determined in terms of network parameters.
\end{abstract}
\vspace*{1cm}
\section{Introduction}

Recently, Rosner \cite{Rosner 1} (see also \cite{Kostelecky-Roberts})
has proposed an interesting analogy between the physics of the
 weak decay of neutral K-mesons (kaons) and a classical system of
 oscillators either electrical or mechanical.
The proposal in \cite{Rosner 1}  makes use of an electric network
composed by two subnetworks of  L$-$C oscillators connected through
a simple quadripole (two-port network).

The analogy in form pointed out in \cite{Rosner 1} is based upon the
observation that the characteristic matrix of the circuit has a
similar aspect to that of the effective Hamiltonian of the neutral
kaon system. Afterwards in \cite{Cocolicchio 1}, an extension to a
general quadripole interaction between L$-$C subnetworks was
presented. The general idea of our work is to formalize the previous
analysis and to extend the character of that observation going
beyond an \textit{analogy}, and to present an \textit{equivalence},
stricto sensu, between the $\mathrm{Schr\ddot{o}dinger}$ dynamics of
a quantum system with a finite number of basis states  \cite{Feynman
1} and a classical dynamics. The equivalence we present is an
isomorphism that connects in univocal way both dynamical systems.

We base our conclusions on the concept of $^\backprime$equivalent dynamics$’$
defined starting from general mathematical concepts \cite{Hirsch},
\cite{Arnold1}. Since the $\mathrm{Schr\ddot{o}dinger}$ dynamics is
defined on a complex space, while the classical dynamics of interest
implies real evolution equations, we profit from the
\textit{decomplexification} concept that allows one to work in a
common space \cite{Hirsch}-\cite{Arnold1}. In this way we are able
to define the general class of classical $\mathcal{C}$ and quantum
$\mathcal{Q}$ systems where the equivalence is valid and
simultaneously, we build up the isomorphism $\Phi_S$, mapping
$\mathcal{C}$ into $\mathcal{Q}$ that includes the general
correspondence between states of both systems. After this formal
development, we treat the particular case of neutral kaons system, denoted by
$\mathcal{K}^\mathrm{o}$ ($|\mathrm{K^o}\rangle$ and
$|\bar{\mathrm{K}}^\mathrm{o}\rangle$), that belongs to the general case of
quantum systems $\mathcal{Q}$, particularly interested in the
aspects of Charge Conjugation-Parity, $\mathrm{CP}$, invariance
\cite{Lee}. In the context of validity of $\mathrm{CPT}$ symmetry
\cite{PDG 1}-\cite{PDG 2}-\cite{PDG 3}, we then consider the equivalent analysis
of Time Reversal, $\mathrm{T}$, invariance.

Just to be self-contained we include a brief summary of the electric
network formalism \cite{Bala} we need. By making use of the concept
of circuital \textit{duality} \cite{Bala}-\cite{Carlin Giordano1},
it was possible to obtain two equivalent electrical representations
of the same classical differential equation, a fact that facilitates
enormously the choice of the parameters that govern the
equivalence of $\mathrm{CP}$ or $\mathrm{T}$ violation in the
network. This class of electric networks $\mathcal{R}$ is univocally
related to the kaon system $\mathcal{K}^\mathrm{o}$ because we find
the complete map between the matrix elements of the effective
Hamiltonian of kaons and those elements of the classical dynamics of
the networks. Moreover, there exists a one to one relationship
between the states $|\mathrm{K^o}\rangle$ and
$|\bar{\mathrm{K}}^\mathrm{o}\rangle$ and port voltages, or
currents, of the electric network.

Following these lines we can give a formal classical test of the
$\mathrm{CP}$ invariance that is a reflection of the quantum test.
From this test, together with the concept of dual network, one
concludes that any violation of the  $\mathrm{CP}$ (or
$\mathrm{T}$) symmetry is  directly related to the non-reciprocity
of the network \cite{Rosner 1}. In fact, the observable associated
to the violation of $\mathrm{T}$ invariance at the quantum level is
associated to the conductance of a non-reciprocal element needed to
be included in the network, the gyrator. The gyrator is a two-port,
non-reciprocal, passive network without losses, and violates the
classical symmetry $\mathrm{T}$ \cite{Bala}-\cite{Carlin
Giordano1}-\cite{Tellegen}. In this way, we end up with a network
completely equivalent to the kaon system, that allows one to present
the relevant parameters of the quantum system in terms of circuit
parameters. The interaction between both L$-$C subnetworks gives
rise to a shift in the proper initial free frequencies, in the same
way as the masses of kaons. Moreover, the presence of proper
relaxation times of the circuit are associated to the mean lives of
K-\textit{short} and K-\textit{long}.

In Section \ref{Dinamicas equivalentes} we analyze the equivalence between quantum and
classical dynamics. Section \ref{Kaones Neutros} is devoted to the neutral kaon system,
while Section \ref{Redes Elec} contains a brief account of electric networks. The
correspondence between those systems is presented in Section \ref{Aplicacion de la correspondencia}. Finally in Section \ref{Conclusiones} we state our conclusions.
\vspace*{1cm}
\section{On the Equivalence Between Dynamics} \label{Dinamicas equivalentes}

It is well known that two systems of equations
$\mathbf{\dot{X}}(t)=\mathbf{AX}(t)$,
$\mathbf{\dot{Y}}(t)=\mathbf{BY}(t)$ are linearly equivalent iff the
corresponding system matrices $\mathbf{A}$ and $\mathbf{B}$ are
related through a similarity transformation, namely
$\mathbf{B}=S^{-1}\mathbf{A}S$ \cite{Arnold1}. We note this
equivalence relation as $\mathbf{A}\sim \mathbf{B}$ and is defined
through  a linear application
$\Phi_S(\mathbf{A})$=$S^{\mathrm{-1}}\mathbf{A}S =\mathbf{B}$. In
summary, fixing a matrix $S$, the application $\Phi_S$ is an
isomorphism.

Consequently, for our purposes of connecting classical and quantum
systems, it is necessary for both of them to be defined in spaces of
finite dimensions. In general, a classical system $\mathcal{C}$ with
$m$ degrees of freedom has its states defined on a real differential
manifold \cite{Marsden1}-\cite{Marsden2} of dimension $m$ whereas a
quantum system $\mathcal{Q}$ with $n$ states belongs to a complex
Hilbert space of dimension $n$. This implies that the relationship
between these systems could be not trivial. In fact, one has to
profit from the so called \textit{decomplexification} procedure
\cite{Arnold1} that we briefly remember here.

To every matrix $\mathbf{A}\in\mathbb{C}^{n\times n}$, the application $ \boldsymbol{\nu}:\mathbb{C}^{\mathrm{\textit{n}x\textit{n}}}\longrightarrow \mathbb{R}^{\mathrm{2\textit{n}\,x\,2\textit{n}}}$ is called matrix decomplexification
\begin{equation*}
\boldsymbol{\nu}(\mathbf{A})=
\left( \begin{smallmatrix}
\Re(\mathrm{A}_{11})     &-\Im(\mathrm{A}_{11})         &\cdots  &\cdots \\
\Im(\mathrm{A}_{11})     &  \Re(\mathrm{A}_{11})        &\cdots  &\cdots\\
\vdots          & \vdots  & \Re(\mathrm{A}_{\textit{nn}})&-\Im(\mathrm{A}_{\textit{nn}})\\
\vdots          & \vdots  & \Im(\mathrm{A}_{\textit{nn}})& \Re(\mathrm{A}_{\textit{nn}})
\end{smallmatrix}\right)
\end{equation*}

Clearly, for the case $n$=1 (a complex number \textit{z}) one has
$ \boldsymbol{\nu}(z)=
\left(
\begin{smallmatrix}
\Re(z)&-\Im(z)\\
\Im(z)&  \Re(z)
\end{smallmatrix}
\right)
$, here $\Re(z)$ and $\Im(z)$ stand for real and imaginary part of complex number $z$ respectively.

This application $\boldsymbol{\nu}$ has several properties that are used below in our analysis.
These properties can be extended to the case of complex matrices $\forall\,M,N,\in \mathbb{C}^{n\times n}$
\begin{align}
 \boldsymbol{\nu}(M+N)&=\boldsymbol{\nu}(M)+ \boldsymbol{\nu}(N)\\
 \boldsymbol{\nu}(M.N)&=\boldsymbol{\nu}(M) \boldsymbol{\nu}(N)\label{dec2}
\end{align}
giving rise to $ \boldsymbol{\nu}([M,N])$=$[\boldsymbol{\nu}(M), \boldsymbol{\nu}(N)]$, then $[M,N]=0\Leftrightarrow[ \boldsymbol{\nu}(M), \boldsymbol{\nu}(N)]=0$. In particular these properties are valid for complex numbers. The symbol 
 $[\bullet,\star]$ is a binary operation named commutator, defined under the set of square matrices.

A way to find the similarity transformation between both systems is to take both matrices
\textbf{A} and \textbf{B} to a diagonal form or in the general case to the Jordan form that always exists.

With this material we are ready to establish the equivalence between classical and quantum systems, the base of our arguments.

\subsection{Classical Systems $\mathcal{C}$}\label{Sist Clasicos}

Given that the superposition principle applies to quantum systems, evolution equations are linear, then the classical systems must be linear. We consider here a system of linear differential equations of second order\label{Equiv Din. clas cuant}
\begin{equation}\label{Sist orden 2 lineal}
\mathbf{A_\mathrm{2}\ddot{q}}_{(t)}+\mathbf{A_\mathrm{1}\dot{q}}_{(t)}
+\mathbf{A_\mathrm{0}q}_{(t)}=\mathbf{0}
\end{equation}
with $\mathbf{q}:\mathbb{R}\longrightarrow\mathbb{R}^{m\times 1}$,  $\mathbf{A}_j\in\mathbb{R}^{m\times m}$, that defines a class $\mathcal{C}$.

We deal in general with systems where \textbf{A}$_2$ has an inverse. Then defining $\mathbf{A}:= \mathbf{A}^{-1}_\mathrm{2}\mathbf{A}_\mathrm{1}$ and $\mathbf{B}:= \mathbf{A}^{-1}_\mathrm{1}\mathbf{A}_\mathrm{0}$ one ends with the equation
\begin{equation}\label{A B}
\mathbf{\ddot{q}}_{(t)}+\mathbf{A\dot{q}}_{(t)}+\mathbf{Bq}_{(t)}=\mathbf{0}  \end{equation}

Clearly, one can reduce the order of the equations by duplicating the number of them.
 This is performed by defining a vector $\textbf{X} \in \mathbb{R}^{\mathrm{2}m}$ such that $\mathbf{X} := (\mathbf{q},\mathbf{\dot{q}})^\intercal$. Then \textbf{X} satisfies
\begin{equation} \label{dcl}
\dot{\mathbf{X}}=\complement\mathbf{X}
\end{equation}
with
\begin{equation}\label{matriz A}
\complement =
\left(\begin{matrix}
\mathbf{0}_m&\mathbf{1}_m\\
\mathbf{-B} &\mathbf{-A}
\end{matrix} \right)
\end{equation}
and $\mathbf{0}_{m},\mathbf{1}_m$ are the zero matrix and the unit matrix of \textit{m}x\textit{m}, respectively.

In this way, the vector \textbf{X} contains all the information of the classical system expressed in the phase space $(\mathbf{q},\mathbf{\dot{q}})$.

\subsection{Quantum Systems $\mathcal{Q}$}\label{Sist CUan}

Let us now consider a quantum system of \textit{n}-base states, each one described by a vector $\psi(t)$ on $\mathbb{C}^n$ that can be written as
$\psi(t)= \textbf{(}\psi_1(t),\hdots,\psi_n(t)\textbf{)}^\intercal$ in terms of the coordinates $\psi_j(t)$=$\langle\textit{j}\:|\varphi(t)\rangle$. This
 $ \psi(t) $ satisfies the  $\mathrm{Schr\ddot{o}dinger}$ equation \cite{Dirac}-\cite{Galindo}
\begin{equation}\label{Sch}
 \dot{\psi}(t)=\mathbf{K}\psi(t)
\end{equation}
where $\mathbf{K} \in\mathbb{C}^{n\mathrm{x}n}$, with elements $\mathbf{K}_{ij}$ = -$\imath\langle|i\mathbf{\hat{H}}|j\rangle$.

Now, in general, given the complex $n$-uple $\boldsymbol{z}$ one defines the  \textit{vector decomplexification} $\mathfrak{D}:\mathbb{C}^{n}\longrightarrow\mathbb{R}^{2n}$, such that
\begin{equation}\label{Decomplex vect}
\mathfrak{D}(\boldsymbol{z})=\textbf{(}\Re(z_1),\Im(z_1),\cdots,\Re(z_n),\Im(z_n)\textbf{)}^\intercal
\end{equation}

Then making a separation between real and imaginary parts of \eqref{Sch} one gets
$d_t[\mathfrak{D}(\psi_{(t)})]= \boldsymbol{\nu}(\mathbf{K})\mathfrak{D}(\psi_{(t)})$, taking by definition
\begin{equation}\label{D-decompl}
\Upsilon(t):=\mathfrak{D}(\psi_{(t)})
\end{equation}
allows to write
\begin{equation} \label{dcu}
\dot{\Upsilon}(t)=\boldsymbol{\nu}(\mathbf{K})\Upsilon(t)
\end{equation}
where clearly $\boldsymbol{\nu}(\mathbf{K})$ is the decomplexification of the matrix $\mathbf{K}$.

\subsection{Construction of the Isomorphism $\Phi_S$}\label{Construcción del iso}

We now build the isomorphism $\Phi_S$ between the dynamics $\mathcal{C}$ and $\mathcal{Q}$ by means of the associated characteristic polynomials
$\complement$ and $\boldsymbol{\nu}(\mathbf{K})$, respectively.

No doubt, a necessary condition for the equivalence we are looking for is that the number of classical degrees of freedom has to be equal to the number of quantum states involved. As we already stated, both systems are linearly equivalent iff $\exists S$ non singular and real such that
\begin{equation}\label{equiv}
\complement = S\: \boldsymbol{\nu}(\mathbf{K})S^{-1}
\end{equation}
$S$ being defined up to a scalar (real or complex), allows one to choose its determinant equal to one.

By using now the general properties of block matrices \cite{Silvester} one computes the polynomial
 $p_{_\complement}$
\begin{equation}
p_{_{\complement}}(z)=det(z\textbf{1}-\complement)= det\left(\begin{smallmatrix}
z\mathbf{1}_n&\mathbf{-1}_n\\
\mathbf{B} &z\mathbf{1}_n+\mathbf{A}
\end{smallmatrix} \right)
\end{equation}
to obtain
\begin{equation}\label{polinomio A}
p_{_{\complement}}(z)=det(z^2\mathbf{1}_n+\mathbf{A}z+\mathbf{B})
\end{equation}

The matrix $z^2\mathbf{1}_n+\mathbf{A}z+\mathbf{B}$ has polynomials of second order in $z$ in the principal diagonal, namely $z^2+\mathbf{A}_{ii}z+\mathbf{B}_{ii}$. The rest of the elements are of first order in $z$, $\mathbf{A}_{ij}z+\mathbf{B}_{ij}$  $\forall i\neq j$.

If \textbf{A} and \textbf{B} are both diagonal, from equation (\ref{polinomio A}) follows that
\begin{equation}\label{polinomio A diag}
p_{_{\complement}}(z)=\prod_{j=1}^n (z^2+\mathrm{A}_{j}z+\mathrm{B}_{j})
\end{equation}
In other words, $p_{_{\complement}}(z)$ is factorized into second degree polynomials.

On the other hand, the polynomial $p_{ \boldsymbol{\nu}(\mathrm{K})}$, by means of the properties of $\boldsymbol{\nu}$ can be written as

\begin{equation}
p_{ \boldsymbol{\nu}(\mathrm{K})}(z)=p_{_\mathrm{K}}(z).p_{_\mathrm{K^{\dagger}}}(z)=\prod_{j=1}^r[z^2-2\alpha_j z+|\lambda_j|^2]^{d_j}=\prod_{j=1}^r[(z-\lambda_j)(z-\lambda^*_j)]^{d_j}\label{polinomio K}
\end{equation}
with $\{\lambda_j,d_j \}$ roots and algebraic multiplicities of $\boldsymbol{\nu(\mathrm{K})}$ and $\alpha=\Re(\lambda_j)$ respectively.

The polynomial associated to the matrix $ \boldsymbol{\nu}(\mathbf{K})$ has the same roots as \textbf{K} plus the corresponding complex conjugate.

If $ \mathsf{Q} $ and $ \mathsf{P} $ are the matrices that transform \textbf{K} and $\complement$ to the Jordan form on $\mathbb{C}$ and $\mathbb{R}$, respectively, the construction of \textit{S} is based precisely on the equality of the Jordan forms $\mathbf{J}_{_{\complement}}=\mathbf{J}_{ \boldsymbol{\nu}(K)}$ that we write
\begin{align}
\mathbf{J}_{_{\complement}}&=\mathsf{P}^{-1}\complement\,\mathsf{P}\label{Jc}\\  \mathbf{J_{{\mathrm{K}}}}&=\mathsf{Q}^{-1}\mathbf{K}\,\mathsf{Q}\label{Jk}
\end{align}
then applying $\boldsymbol{\nu}$ to the last expression and using the property \eqref{dec2} one has $ \boldsymbol{\nu}(\mathbf{J_{{\mathrm{K}}}})=\boldsymbol{\nu}(\mathsf{Q})^{-1} \boldsymbol{\nu}(\mathbf{K})\, \boldsymbol{\nu}(\mathsf{Q})$ and finally $ \mathbf{J}_{ \boldsymbol{\nu}(\mathrm{K})}= \boldsymbol{\nu}(\mathbf{J_{{\mathrm{K}}}})$ \cite{Caruso}.

Being assumed that $p_{ \boldsymbol{\nu}({\mathrm{K}})}=p_{_{\complement}}(z)$, one has $\mathbf{J}_{_{\complement}}=\mathbf{J}_{ \boldsymbol{\nu}(K)}$:

\begin{equation}\label{eq}
\complement=\mathsf{P} \boldsymbol{\nu}(\mathsf{Q})^{-1} \boldsymbol{\nu}(\mathbf{K})\,[\mathsf{P} \boldsymbol{\nu}(\mathsf{Q})^{-1}]^{-1}
\end{equation}

From equation \eqref{equiv} the desired matrix $S$ appears defined by
\begin{equation}\label{S:}
S:=\mathsf{P}\: \boldsymbol{\nu}(\mathsf{Q})^{-1}
\end{equation}

We would like to remark that when the equivalence between the dynamics driven by $\complement$ and $ \boldsymbol{\nu}(\mathbf{K})$ exists, i.e. \:$p_{_{\complement}}$=$p_{ \boldsymbol{\nu}(\mathbf{K})}$ and assuming that each eigenvalue of \textbf{K} is not degenerate ($d_j=1$,  $j=1,...,r$) and that \textbf{A} and \textbf{B} are digonalized in the same basis, the identification of polynomials is immediate
\begin{align}
\mathrm{A}_{j}&=-2\Re\:(\lambda_j)\label{relacion de autov1}\\
\mathrm{B}_{j}&=|\lambda_j|^2 \label{relacion de autov2}
\end{align}
and consequently $4\mathrm{B}_{j}>\mathrm{A}_{j}^2$.

In summary, using the definition $\mathbf{K:=-\imath H}$, the process driven the equivalence between systems $\mathcal{C-Q}$ can be expressed as
\begin{equation*}
\begin{matrix}
\psi\in\mathbb{C}^n,\: \dot{\psi}(t)=\mathbf{K}\psi(t)&\qquad \qquad & \ddot{q}+\mathbf{A}\dot{q}+\mathbf{B}q=0,\: q\in\mathbb{R}^n\\
\Bigg\downarrow\vcenter{%
\rlap{$ \mathfrak{D}$}}&\qquad  \qquad & \Bigg\updownarrow\vcenter{%
\rlap{$\mathbf{x}=(q,\dot{q})^\intercal$}}\\
\Upsilon\in\mathbb{R}^{2n}, \:\dot{\Upsilon}(t)= \boldsymbol{\nu}(\mathbf{K})\Upsilon(t)&
 \boldsymbol{\underleftrightarrow{ \qquad \boldsymbol{\nu}(\mathbf{K})\sim \complement\qquad }} &\dot{\mathbf{X}}(t)=\complement \mathbf{X}(t), \:\mathbf{X}\in\mathbb{R}^{2n}
\end{matrix}
\end{equation*}

The transformation from $ \boldsymbol{\nu}(\mathbf{K})$ to $\complement$ is done by means of $\boldsymbol{\Phi}_S$, with $S$ non singular
such that:
$\boldsymbol{\Phi}_{S}\footnotesize{\mathrm{(\boldsymbol{\complement})}}$=$\boldsymbol{ \boldsymbol{\nu}}_{(\mathbf{K})}$. Clearly, the application $\boldsymbol{\Phi}_S$ is an isomorphism between the quantum systems $\mathcal{Q}$ and the classical ones $\mathcal{C}$ $\boldsymbol{\Phi}_S:\mathcal{C}\longleftrightarrow\mathcal{Q}
$, because both classes are characterized by  $\boldsymbol{\nu(\mathbf{K})}$ and $\complement$ respectively.

\vspace*{1cm}
\section{Quantum Dynamics of Neutral Kaons}\label{Dina S Granticos}\label{Kaones Neutros}

We are particularly interested in the quantum system of neutral kaons,
denoted by $\mathcal{K}^\mathrm{o}$, because, under the hypothesis of
Wigner-Weisskopf \cite{Wigner-Weisskopf}, it can be written as a two-state system. This allows one to exemplify very easily the equivalence with a classical system. Moreover, we want to give a formal context to the Rosner \cite{Rosner 1} and  Cocolicchio \cite{Cocolicchio 1} proposals to establish  an analogy of the kaon system with electric networks.

General principles on the basis of the Quantum Field Theory guarantee the validity of the CPT symmetry in Nature \cite {Froggatt Nielsen}-\cite{Streater Wightman}-\cite {Luders}-\cite{Sachs}. When CPT is a symmetry, the mass of a particle and its antiparticle has to be equal
\cite{Lee}. This equality is experimentally verified in the case of neutral kaons in an impressive way \cite{PDG 1}-\cite{PDG 2}-\cite{PDG 3}:
\begin{equation}
\frac{|m_{\mathrm{K}^{\circ}}-m_{\bar{\mathrm{K}}^{\circ}}|}{m_{\mathrm{K}^{\circ}}+m_{\bar{\mathrm{K}}^{\circ}}} <4\times10^{-19}
\end{equation}
Consequently, in this context, it is equivalent to speak about CP or T invariance, or non-invariance.

We consider here the weak decay of the neutral kaons $\mathrm{K}^{\circ}$,   $\mathrm{\bar{K}}^{\circ}$ in the standard formalism \cite{Lee Oehme Loy}
that includes the strangeness selection rule $\Delta$S=$\pm$1.

All the elements of the Hilbert space $\mathcal{H}$ are eigenstates of  $\mathbf{\hat{H}_o}=\mathbf{\hat{H}_s}+\mathbf{\hat{H}_\gamma}$ where $\mathbf{\hat{H}_s}$ and $\mathbf{\hat{H}_\gamma}$ stand for the strong and electromagnetic interactions respectively. To this one adds the small perturbation of the weak interaction $ \mathbf{\hat{H}}_w $, that gives rise to transitions among the states of $\mathbf{\hat{H}_o}$.

In order to formalize the quantum dynamics we split the Hilbert space into the direct sum $\mathcal{H}=\mathcal{A}\oplus\mathcal{B}$, where $\mathcal{A}$ corresponds to the states of $\mathrm{K}^{\circ}$ and $\mathrm{\bar{K}}^{\circ}$ while $\mathcal{B}$ includes all the possible decay final states. We call $\{|j\rangle\}$ and $\{|\beta\rangle\}$ orthogonal complete basis of $\mathcal{A}$ and $\mathcal{B}$ respectively.

Consequently, any state at time $t$ is represented by
\begin{equation}|\Psi(t)\rangle =\sum_j | j \rangle \langle j|\Psi(t)\rangle+\underset{\beta}{\text{\huge{S}}}|\beta \rangle \langle \beta|\Psi(t)\rangle\;
\end{equation}
where $ \underset{\beta}{\text{\LARGE{S}}} $ is a sum on $\mathcal{B}$ states.

Defining now $a_j(t):=\langle j|\Psi(t)\rangle$, with $j$=1,2 corresponding to $\mathrm{K}^\circ$ and $\bar{\mathrm{K}}^\circ$ respectively, and $c_\beta(t):=\langle \beta|\Psi(t)\rangle$ one can describe the subdynamics on $\mathcal{A}$
by the coordinates $a_j(t)$ or, equivalently, by considering the one to one correspondence $|\mathrm{K}^\circ\rangle\longleftrightarrow(1,0)^\intercal$, $|\bar{\mathrm{K}}^\circ\rangle\longleftrightarrow(0,1)^\intercal$, the $\mathbb{C}^2$
vector
\begin{equation}\label{Vector en A}
\psi(t)=\textbf{(}a_1(t),a_2(t)\textbf{)}^\intercal
\end{equation}
also describes the same dynamics.

Just to complete the description it is necessary to give the initial condition for the evolution that is written as $a_j(0)\neq 0$, $c_\beta(0)=0$, or in other words $|\Psi$(0)$\rangle \in\mathcal{A}$.

The evolution equation of the dynamics under consideration described by $\psi(t)$ is \cite{Lee}:
\begin{equation}\label{Dinamica en A}
\imath d_t\psi(t)=\mathbf{(M-\imath\Gamma)}\psi(t)
\end{equation}
where $\mathbf{M}$ and $\mathbf{\Gamma}$ are hermitian matrices.
Clearly, the matrix $\boldsymbol{\Gamma}$ takes into account the decay width.

All the information on the decay channels is contained in \eqref{Dinamica en A} as is clear from the matrix elements of
 $\mathbf{M-\imath\Gamma}$:
\begin{align}\label{M y Gamma}
\mathrm{M}_{ij}=\langle i|\mathbf{\hat{H}}|j\rangle +\underset{\beta}{\text{\huge{S}}}\frac{\langle i|\mathbf{\hat{H}}_w|\beta\rangle\langle\beta|\mathbf{\hat{H}}_w|j\rangle}{E_j-E_\beta},\quad
\mathrm{\Gamma}_{ij}=\pi \underset{\beta}{\text{\huge{S}}} \langle i|\mathbf{\hat{H}}_w|\beta\rangle\langle\beta|\mathbf{\hat{H}}_w|j\rangle \:\delta(E_j-E_\beta)
\end{align}
with $E_j$ and $E_\beta$ being the eigenvalues of  $\mathbf{\hat{H}_o}$ associated to the basis states.

When the Hamiltonian $\mathbf{\hat{H}}$ has a given symmetry, there exist relations among the matrix elements  $\mathbf{H=M-\imath \Gamma}$.

In \cite{Lee} we see that if CPT is a symmetry, one has $\mathrm{M}_{11}=\mathrm{M}_{22}$ and $\mathrm{\Gamma}_{11}=\mathrm{\Gamma}_{22}$, while if T is also a symmetry, then
$\mathrm{M}_{12}=\mathrm{M}_{21}$ and $\mathrm{\Gamma}_{12}=\mathrm{\Gamma}_{21}$.

In other words, the symmetry, when projected in the subspace $\mathcal{A}$, provides relationships among the matrix elements of $\mathbf{H}$
and consequently provides a \textit{test of violation}.

Before the crucial experiment \cite{Christenson} everything pointed to $\mathbf{CP}$ as a good symmetry. This implied that,
with the election of the phases of the states $\{|\mathrm{K^o}\rangle,|\mathrm{\bar{K}^o}\rangle\}$ in such a way that $\mathbf{\hat{C}\hat{P}}|\mathrm{K^o}\rangle =|\mathrm{\bar{K}^o}\rangle$ \cite{Sachs}, one can write states with well defined CP, given
naturally as the mixtures $|\mathrm{K_1}\rangle =\tfrac{|\mathrm{K^o}\rangle+|\mathrm{\bar{K}^o}\rangle}{\sqrt{2}},\: |\mathrm{K_2} \rangle =\tfrac{|\mathrm{K^o}\rangle-|\mathrm{\bar{K}^o}\rangle}{\sqrt{2}}$, obtained through the change of basis given by the unitary matrix
\begin{equation}\label{Q CP inv}
\mathfrak{Q}=\tfrac{1}{\sqrt{2}}\left(\begin{matrix}
1&\,1\\
1&\,-1
\end{matrix}\right)
\end{equation}
that diagonalizes the Hamiltonian $\mathbf{H=M-\imath\Gamma}$.

CP being conserved, the only decays into pions allowed are
$|\mathrm{K_1}\rangle \longmapsto |2\pi\rangle, \:
|\mathrm{K_2}\rangle \longmapsto |3\pi\rangle$
because the states $|2\pi\rangle$ and $|3\pi\rangle$ have well defined CP, namely $\mathbf{\hat{C}\hat{P}}|2\pi\rangle=|2\pi\rangle$ and $\mathbf{\hat{C}\hat{P}}|3\pi\rangle=-|3\pi\rangle$ \cite{Lee}-\cite{Sakurai}. These two decay modes have certainly very different
 decay times. In fact, the first one has $\tau$ of the order of $10^{-10} s$ while the second is of the order of $5\times 10^{-8} s$.

\subsection{CP Violation}\label{Violacion de CP}

 CP violation was observed for the first time by Christenson, Cronin, Fitch and Turlay  \cite{Christenson}. They experimentally found that $|\mathrm{K_2}\rangle$ is able to decay also into two pions. This unexpected fact shows that the weak interactions violates CP symmetry. Consequently, the matrix $\mathbf{M-\imath\Gamma}$ now is diagonalized by means of a new matrix  $\mathsf{Q}$, that takes into account the mixture of CP states and that can be written as
\begin{equation}
\mathsf{Q}(\epsilon)=\frac{1}{\sqrt{2(1+|\epsilon|^2)}}\left(\begin{matrix}
1+\epsilon &\quad1+\epsilon\\
1-\epsilon &-(1-\epsilon)
\end{matrix}\right):=(\boldsymbol{\mathsf{S}},\boldsymbol{\mathsf{L}})\label{Q CP NO inv}
\end{equation}

In this way, the eigenstates of $\mathbf{M-\imath\Gamma}$ expressed in the ket basis \{$|\mathrm{K^o}\rangle,|\mathrm{\bar{K}^o}\rangle$\} are now
\begin{align}
|\mathrm{K_\text{\tiny{S}}^o}\rangle=\dfrac{1}{\sqrt{2(1+|\epsilon|^2)}}
[(1+\epsilon)|\mathrm{K^o}\rangle+(1-\epsilon)|\mathrm{\bar{K}^o}\rangle],\quad|\mathrm{K_\text{\tiny{L}}^o}\rangle=\dfrac{1}{\sqrt{2(1+|\epsilon|^2)}}
[(1+\epsilon)|\mathrm{K^o}\rangle-(1-\epsilon)|\mathrm{\bar{K}^o}\rangle]
\end{align}
where, as usual, the indices \begin{scriptsize}S, L\end{scriptsize} are realated to the decay times \textit{short, long} respectively.

It can be shown \cite{Caruso} that when CPT is valid, the condition $\mathrm{H}_{12}=\mathrm{H}_{21}$ is equivalent to $\mathsf{Q}$ being unitary. Then, clearly, the complex
parameter $\epsilon$ that makes the matrix $\mathsf{Q}$ non unitary, is a measure of the amount of CP violation. Note that  $\mathsf{Q}(\epsilon)$ is unitary if $\epsilon=0$. The parameter $\epsilon$ can be written in terms of the matrix elements of \textbf{H} as
\begin{equation}
\epsilon = \frac{\sqrt{{\mathrm{H}_{12}}}-\sqrt{{\mathrm{H}_{21}}}} {\sqrt{{\mathrm{H}_{12}}}+\sqrt{{\mathrm{H}_{21}}}}\label{Epsilon posta}
\end{equation}

Notice naming $\boldsymbol{\mathsf{S}}$ and $\boldsymbol{\mathsf{L}}$ the first and second column of $\mathsf{Q}$, respectively, see \eqref{Q CP NO inv}, and denoting by $\boldsymbol{.}$ the usual scalar product, or \textit{dot product}, of vectors, the test of CP violation can be enunciated as: \textit{if $\boldsymbol{\mathsf{S}.\mathsf{L}}\neq 0$ then, CP is not conserved}.

It is precisely this new expression of the test of CP violation that can be easily translated to the equivalent classical dynamical system via the isomorphism $\boldsymbol{\Phi}_S$.

\vspace*{1cm}
\section{Electric Networks ($\boldsymbol{\mathcal{R}}$)}\label{Redes Elec}

An electric network \cite{Bala}-\cite{Carlin Giordano1} includes a set of elements together with a given way of connections among them. These elements can be classified into five classes, namely: resistors \{R\}, capacitors \{C\}, inductances \{L\}, voltage generators \{$v_g$\} and current generators \{$i_g$\}.
We are particularly interested in localized circuits where voltage and current depend only upon time. The standard linear relations among the classes are
{\setlength\arraycolsep{1pt}
\begin{eqnarray}\label{RCL}
v_\mathrm{R}(t)=\mathrm{R}\,i_\mathrm{R}(t),\quad
i_\mathrm{C}(t)=\mathrm{C}\: d_t v_\mathrm{C}(t),\quad
v_\mathrm{L}(t)=\mathrm{L}\: d_t i_\mathrm{L}(t)
\end{eqnarray}

The way that the elements are connected is determined by a \textit{linear graph}, defined by nodes connected by edges. When the edges have a given sense, provided by the current flowing through it, one says that the graph is \textit{oriented}, and each edge of the graph is called arc. The oriented graph $\mathcal{G}$ is noted by means of a cartesian product of the sets of nodes and arcs $\mathcal{G=(N,A)}$. We are interested in \textit{connected} graphs, those where there is at least one path of $\mathcal{G}$ between any pair of nodes of $\mathcal{G}$.

The corresponding dynamics of an electric network is defined by the appropriate use of the relations ~\eqref{RCL} together with the Kirchhoff rules that take care of the topology of the network.

The network has ports: pairs of terminals that allow exchange of energy with the surrounding. We restrict our analysis to passive networks, where the energy provided by an external source is non-negative. In any case one has the possibility of choosing the voltage or the current as the representative variable of the excitation or of the response of the network.

We call \textbf{v}$(t)$ the vector corresponding to the port voltage and \textbf{V}$(s)$ its Laplace transform. A similar notation is used for currents, \textbf{i}$(t)$ and \textbf{I}$(s)$ respectively. The network, initially without external energy, provides the relation
\begin{equation}\label{M-represent}
\boldsymbol{\mathcal{O}}(s)=\boldsymbol{\mathcal{M}}(s)\boldsymbol{\mathcal{J}}(s)
\end{equation}
between the Laplace transform of the output $\boldsymbol{\mathcal{O}}$ and the input $\boldsymbol{\mathcal{J}}$ \cite{Bala}. Consequently, $\boldsymbol{\mathcal{M}}$ is the network matrix.

From $\mathcal{O}_{\alpha}(s)=\sum_{\beta=1}^n \mathcal{M}_{\alpha\beta}(s)\mathcal{J}_{\beta}(s)$ there results
\begin{equation}
\mathcal{M}_{\alpha\beta}(s)=\frac{\mathcal{O}_{\alpha}(s)}{\mathcal{J}_{\beta}(s)}|_{\mathcal{J}_k=0} \,\,\,\forall k \neq \beta
\end{equation}
given rise to the definition of the elements of
\begin{equation}\label{Matriz M}
\mathcal{M}_{\alpha\beta}:=\left.\frac{\mathcal{L}(r_\alpha)}{\mathcal{L}(e_\beta)}\right.
\end{equation}
where $r_\alpha$ and $e_\beta$ stand for response and excitation, respectively.
One then says that the network $\mathcal{R}$ has a $\mathcal{M}$-\textit{representation} or that it belongs to the class $\mathcal{M}$.

When the port voltages can be considered as input and the port currents as output, the corresponding representation of the network is called of \textit{admittance} or type $\mathbf{Y}$, and the input-output relationship is
\begin{equation}\label{Admitance Rep}
\mathbf{I}(s)=\mathbf{Y}(s)\mathbf{V}(s)
\end{equation}
On the other hand, if the port currents are chosen as input and the port voltages as output, one is in the case of the \textit{impedance}
representation or type $\mathbf{Z}$. In this case one has
\begin{equation}\label{Impedance Rep}
\mathbf{V}(s)=\mathbf{Z}(s)\mathbf{I}(s)
\end{equation}
Any other representation is called \textit{hybrid}.

A very important concept, relevant to our discussion is that of \textit{reciprocity}. A network not connected to external energy sources is reciprocal iff considering two different terminals $\alpha$ $\neq$ $\beta$, the excitation in $\alpha$ gives rise to a response in $\beta$
that is invariant under the permutation $\alpha\longleftrightarrow\beta$ or $e_\alpha(t)=e_\beta(t)\Longrightarrow r_\alpha(t)=r_\beta(t)$.

Consequently, $\mathcal{R}$ is \textit{reciprocal} iff $\mathcal{M}_{\alpha \beta}(s)= \mathcal{M}_{\beta\alpha}(s)$ with  $\alpha$ $\neq$ $\beta$.

 The diagonal elements \{$\mathcal{M}_{\alpha\alpha}$\} are called of \textit{input} ones, because they relate the quantities of port $\alpha$, while the elements not belonging to the diagonal \{$\mathcal{M}_{\alpha\beta}$\}, with $\alpha\neq\beta$ are called of \textit{transference} ones, and relate quantities of different ports.

\subsection{Dual Networks}\label{dual network}

When the graph $\mathcal{G}$ associated to the network $\mathcal{R}$ is \textit{planar}, there exists its \textit{dual} graph \cite{Bala} denoted by  $\mathcal{G}^*$. This dual graph corresponds naturally to a \textit{dual network} noted $\mathcal{R}^*$. Consequently, each arc voltage $v_j$ of $\mathcal{R}^*$ changes into a arc current $i_j$ of $\mathcal{R}$ and vice versa.

Clearly, the equations ~\eqref{RCL} have the dual version, namely
\begin{eqnarray*}
\left.\begin{matrix}
v(t)&=&\mathrm{R}\,i(t)\\
i(t)&=&\mathrm{C}\:\mathrm{d}_{t}\,v(t)\\
v(t)&=&\mathrm{L}\:\mathrm{d}_{t}\,i(t)
\end{matrix}\right\rbrace \longleftrightarrow \left\lbrace\begin{matrix}
i(t)&=&\mathrm{G}\,v(t)\\
v(t)&=&\mathrm{L}\:\mathrm{d}_{t}\,i(t)\\
i(t)&=&\mathrm{C}\:\mathrm{d}_{t}\,v(t)
\end{matrix}\right.
\end{eqnarray*}

Or in other words, the dual transformation implies the interchange
\begin{equation}
v\longleftrightarrow i,\quad \text{C}\longleftrightarrow \text{L},\quad
\text{R}\longleftrightarrow \text{G}
\end{equation}

In summary, the connected networks allow for two representations, according to the selection of voltage, or current, port to present
the analysis. The corresponding networks are mutually dual.

\vspace*{1cm}
\section{Correspondence Electric Network - Kaons:  $\boldsymbol{\mathcal{R-\mathcal{K}^\mathrm{o}}}$}\label{Aplicacion de la correspondencia}

The main goal of our proposal is to build an electric circuit able to describe the system of neutral kaons, satisfying the equivalence $\thicksim$ previously introduced. As a result, the matrix elements of the effective Hamiltonian $\mathbf{M-\imath\Gamma}$ are related, by means of a similitude transformation to those of the appropriate electric circuit as
 \begin{equation}
 \complement=-S \boldsymbol{\nu}_{(\mathbf{\Gamma+\imath M})}S^{-1}
 \end{equation}
In this way, the symmetries present in the kaon system and the corresponding tests of validity, have an unique reflection in the electric circuit.

The method of analysis of the time evolution of electric circuits previously introduced, based on nodes, arcs and loops \cite{Bala} ends on a system of linear differential equations of second order with constant coefficients, or alternatively on systems of linear integro-differential equations. In our case, as the precise electric network is not known, one has to make a \textit{synthesis} \cite{Bala} of the set of all electric networks in a given family or subset of this.

The general scheme of the procedure is to look for the network that is equivalent to a quantum system of  \textit{n}-states. This network should have
a first order equivalent equation of the form \eqref{Sist orden 2 lineal}, where the generalized coordinates $\{q_j\}$  are now voltages and currents and with the $\complement$ matrix having  \textit{n} eigenvalues in the third quadrant of the complex plane, together with their complex conjugates.

 One ends with a general structure composed by \textit{n}-subnetwork (dipoles) \{$\mathcal{R}_j$\}$_{j=1,...,n}$ connected among themselves by means of the so called interaction network $\mathcal{R_I}$ of \textit{n} ports \cite{Bala}. Each subnetwork introduces a degree of freedom described by a port voltage, or port current. These time functions provide the entire information of the state $\mathcal{R}_j$ at a given time. Clearly, on each dipole one gives the initial condition. Figure \ref{Red de n-puertos} shows the resulting general topological array.
\begin{figure}[!h]
\centerline{\includegraphics[totalheight=3cm]{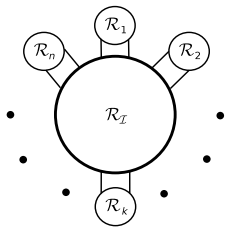}}
\caption{\footnotesize{General electric network $\mathcal{R}$. \label{Red de n-puertos}}}
\end{figure}

Notice that in our case, the network dynamics must be time translation invariant, as required by \eqref{Sist orden 2 lineal}, to be in agreement with
the corresponding property of the quantum system.

In general, a network of \textit{n} ports allows $(2n)! / (n!)^2$ representations. However, only two of them are useful for our purposes, namely those where the mix of state variables is not present: all port variables are either voltages or currents. Consequently, the equivalent network families of interest are type \textbf{Y} or type \textbf{Z}.

The electric network $\mathcal{R}$ of interest in connection with the neutral kaon system should have a classical matrix $\complement$ of $4\times 4$
with two different complex eigenvalues placed on the third quadrant. Then the general topology reduces to the one composed by two subnetworks $\mathcal{R}_1$ and $\mathcal{R}_2$ connected by means of a two-port network $\mathcal{R_I}$, Fig. \ref{Topologia2-puertas}.
\begin{figure}[!h]
\centerline{\includegraphics[totalheight=2.5cm]{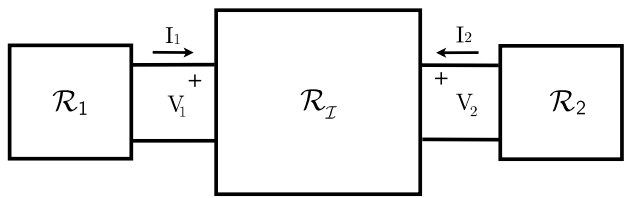}}
\caption{\footnotesize{A network $\mathcal{R}$ composed by two subnetworks (dipoles) $\mathcal{R}_1$ and $\mathcal{R}_2$ connected through $\mathcal{R_I}$.\label{Topologia2-puertas}}}
\end{figure}

In order to fix the structure of each $\mathcal{R}_j$ one starts by observing that the free states (when $\mathcal{R_I}$ is not connected) correspond to $\mathbf{Y}(s)=\mathbf{0}$ or to $\mathbf{Z}(s)=\mathbf{0}$ according to the equations \eqref{Admitance Rep} and \eqref{Impedance Rep}, respectively.
Correspondingly, the free states of the quantum system of kaons are those that evolve with $\mathbf{\hat{H}}=\mathbf{\hat{H}_o}$ and have the form
 $\psi_{\circ j}(t)=\mathrm{A}_j e^{-\imath m_jt}$ of frequency $m_j$. This condition implies that each $\mathcal{R}_j$ should be \textit{reactive}, i.e., without resistive losses, contributing with one proper frequency each. This condition forces one, and only one, connection of inductors and capacitors, depending on the use of voltage or current as the input.
In Fig. \ref{L//C} the possible subnetworks $\mathcal{R}_j$,  $j=1,2$ that should be connected to the ports 1 and 2 of $\mathcal{R_I}$ are shown. Both   dipoles are certainly oscillators with the corresponding frequencies dictated by the values of L$_j$ and C$_j$.
\begin{figure}[!hd]
\centerline{\includegraphics[totalheight=2.5cm]{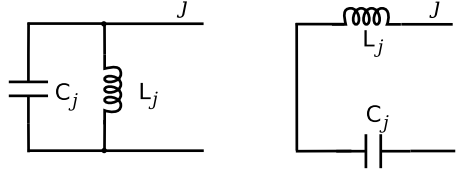}}
\caption{\footnotesize{Reactive subnetworks $\mathcal{R}_j$. Left with $\mathbf{v}(t)$ excitation. Right with $\mathbf{i}(t)$ excitation. \label{L//C}}}
\end{figure}

It is worth noticing that if a network $\mathcal{R}$ having a dual representation $\mathcal{R}^*$, presents an analogous behavior (equivalent according to $\boldsymbol{\Phi}_S$) to a quantum system $\mathcal{Q}$ then, $\mathcal{R}^*$ shares this analogous behavior to $\mathcal{Q}$, i.e.
\begin{equation*}
\forall \mathcal{R}: \exists \mathcal{R}^*\big/ \Big\{ \mathcal{R}\thicksim\mathcal{Q}\Leftrightarrow\mathcal{R}^*\thicksim\mathcal{Q}\Big\}
\end{equation*}
that follows from the property $\mathcal{R}\thicksim\mathcal{R}^*$ and the transitivity of the equivalence $\thicksim$.

Our analysis uses port voltages as excitation variables. Then the complete network can be schematized as in Fig. \ref{Esquema completo de la Red eligiendo a v}
\begin{figure}[!h]
\centerline{\includegraphics[totalheight=2.7cm]{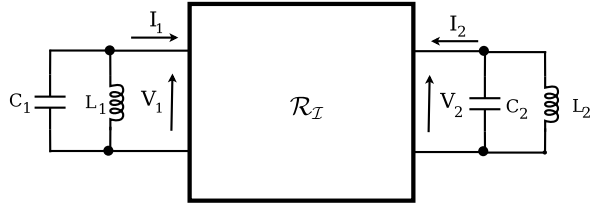}}
\caption{\footnotesize{Complete scheme of the network $\mathcal{R}$ with $\mathbf{v}(t)$ excitation.}\label{Esquema completo de la Red eligiendo a v}}
\end{figure}

We go then to the synthesis of the network $\mathcal{R_I}$ that ends with a family $\mathcal{C}$, equivalent to the kaon system $\mathcal{K}^\text{o}$. We consider connected networks because they have a dual representation.

From Fig. \ref{Esquema completo de la Red eligiendo a v} one has
\begin{equation}\label{kir nodo j}
i_j(t)+i_{\mathrm{C}_j}(t)+i_{\mathrm{L}_j}(t)=0
\end{equation}
Using  now the standard relations between voltages and currents and performing a Laplace transformation one ends with
\begin{equation}\label{V-I tanque}
-\mathrm{I}_j(s)=\mathcal{Y}_j(s){\mathrm{V}_j(s)}-\mathrm{F}_j(s)
\end{equation}
If $t=0$ is the initial time, $s=\sigma+\imath\omega$, F$_j(s)$=C$_j v_j(0)-\frac{1}{s}i_{\mathrm{L}_j}(0)$ includes the initial conditions. We call $\mathcal{Y}_j(s)$ the \textit{admittance} function of the subnetwork $\mathcal{R}_j$, given by
\begin{equation}\label{admitancia//}
\mathcal{Y}_j(s)=\left.\mathrm{C}_js+\frac{1}{\mathrm{L}_js}\right.
\end{equation}

Clearly, the frequency of each subnetwork, $\omega_{oj}$, is given by $\omega_{oj}=1/\sqrt{\mathrm{L}_j\mathrm{C}_j}$

On the other hand, the interaction network $\mathcal{R_I}$, initially without energy sources, is governed by \eqref{Admitance Rep}, with $\mathbf{Y}:\mathbb{C}$ $\longmapsto$ $\mathbb{C}^{2\mathrm{x}2}$, that we write
\begin{equation}
\mathbf{Y}(s)=\left(\begin{matrix}
y_{11}(s)&\: &y_{12}(s)\\
y_{21}(s)&\: &y_{22}(s)
\end{matrix}\right)
\end{equation}
with $y_{ij}(s)=\left. \mathrm{I}_i(s)/\mathrm{V}_j(s)\right|_{V_k=0}$; $j$\,$\neq$\,$k$, called the admittance parameters of the circuit.

Coming back to \eqref{V-I tanque}, it can be written as
\begin{equation}\label{dinamica circuital ADMITACNCIA en s}
s\mathbf{V}(s)+\left(\frac{1}{s}\boldsymbol{\mho}^2+\mathbf{DY}(s)\right)\mathbf{V}(s)=\mathbf{D}\mathbf{F}(s)
\end{equation}
with the matrices \textbf{D}$:=diag($C$_1^{-1}$,C$_2^{-1}$) and $\mathbf{\boldsymbol{\mho}}:=diag(\omega_{o1},\omega_{o2}$).

As was stated before, the states of the free network, $\mathbf{Y}(s)=\mathbf{0}$, are related to the unperturbed states of the quantum system, $\mathbf{\hat{H}}=\mathbf{\hat{H}}_o$. Consequently, the eigenfrequency of each subnetwork correspond to the neutral kaon masses $\omega_{oj}\longleftrightarrow m_j$, with $j$=$\mathrm{\scriptstyle K}^\circ$, $\mathrm{\bar{\scriptstyle K}}^\circ$. Then, CPT invariance implies $\omega_{oj}=\omega_{o}$ for all $j$ that forces
\begin{equation}
\mathbf{\boldsymbol{\mho}}=\omega_o \mathbf{1}
\end{equation}
with $\mathbf{1}$ the identity  $2\times 2$ matrix.

In order to obtain a linear differential equation of second order in time, one needs the initial conditions in \textbf{F}$(s)$. We consider zero port currents, $\mathbf{i}(0)=\mathbf{0}$. Notice that this initial condition is a necessary condition once one considers that $\mathcal{R_I}$ is initially without energy. Therefore can be used the usual methods of networks synthesis.

Then, from \eqref{kir nodo j} to each node and \eqref{RCL} one has
\begin{equation}\label{Corriente inductor}
\mathbf{i}_\mathrm{L}(0)=-\mathbf{D}^{-1}\dot{\mathbf{v}}(0)
\end{equation}
that allows the term $\mathbf{D}\mathbf{F}(s)$ of \eqref{dinamica circuital ADMITACNCIA en s} to be written as
\begin{equation}\label{DF}
\mathbf{DF}(s)=\mathbf{v}(0)+\frac{1}{s}\dot{\mathbf{v}}(0)
\end{equation}

Next we deal with the general form of the matrix \textbf{Y}$(s)$.

We are considering first reciprocal networks that implies that  $y_{ij}(s)=y_{ji}(s), \forall s$. This is the case of any network composed by R, L and C \cite{Bala}. Moreover, a connected reciprocal network of two ports reduces always to the so called structure  $\boldsymbol{\Pi}$ (or its dual \textbf{T}) as is shown in Fig. \ref{PI-T}
\begin{figure}[!h]
\label{PI-T}
\centerline{\includegraphics[totalheight=3cm]{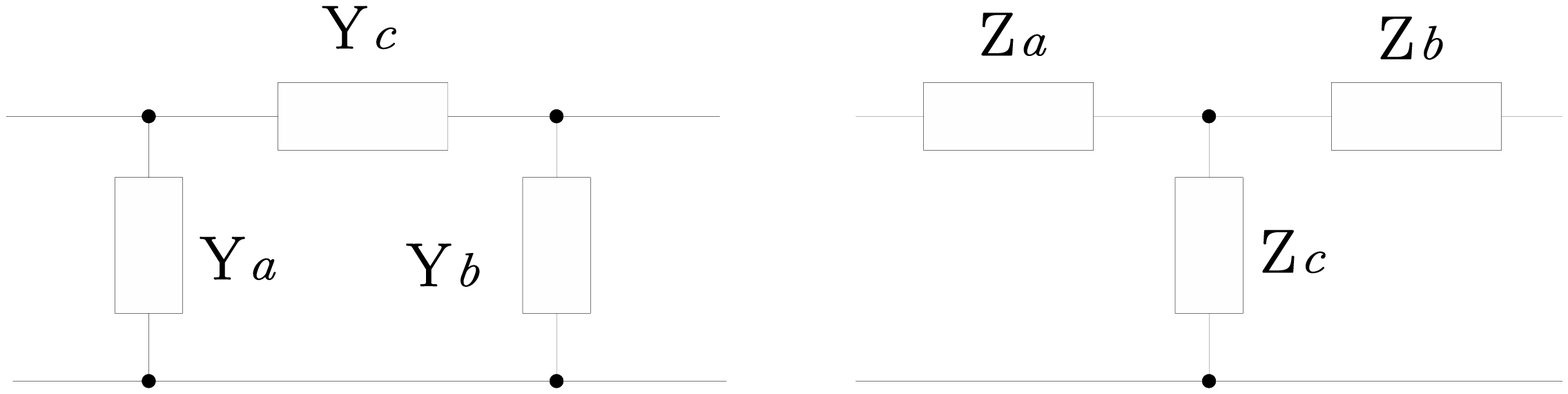}}
\caption{\footnotesize{\textit{Left}: Network with structure $\boldsymbol{\Pi}$. \textit{Right}: Network with structure \textbf{T}}. Both structures are mutually dual.}
\end{figure}

The topology $\boldsymbol{\Pi}$, implies a relationship between \{Y$_a$,Y$_b$,Y$_c$\} and the matrix \textbf{Y} \cite{Bala}
\begin{eqnarray*}
y_{11}=\mathrm{Y}_a + \mathrm{Y}_c,\quad
y_{22}=\mathrm{Y}_b + \mathrm{Y}_c,\quad
y_{12}=y_{21}=-\mathrm{Y}_c
\end{eqnarray*}

We consider first the most simple case of a coupling between subnetworks defined by a matrix \textbf{Y}$(s)$ constant in $s$. This is the case of a matrix $\mathcal{R_I}$ of only resistors. Then
\begin{equation}\label{Y resist}
\mathbf{Y}=\left(\begin{matrix}
\mathrm{G}_a+\mathrm{G}_c&-\mathrm{G}_c\\
-\mathrm{G}_c&\mathrm{G}_b+\mathrm{G}_c
\end{matrix}\right) \end{equation}
where \{$\mathrm{G}_a,\mathrm{G}_b,\mathrm{G}_c$\} are the conductances of $\mathcal{R_I}$.

The complete network results from the connection of a dipole to each port of $\mathcal{R_I}$ whose dynamics is given by
\begin{equation*}
\ddot{\mathbf{v}}(t)=-\boldsymbol{\mho}^2 \mathbf{v}(\tau)-\mathbf{DY\dot{v}}(t)
\end{equation*}

Here one can identify the matrices \textbf{A} and \textbf{B} of Eq. \eqref{A B}, i.e.
\textbf{A=DY}, \textbf{B=$\boldsymbol{\mho}^2$}.
Both matrices can be diagonalized under the same change of basis. The matrix \textbf{DY} is symmetric  having two real positive eigenvalues. For this reason, the real part of the eigenvalues of \textbf{K}, that measure the time life of the decaying particles can be exactly fitted from Eq. \eqref{relacion de autov1}.

Due to the fact that the quantum system of interest presents two different eigenvalues, the network we are looking for has to have a matrix \textbf{B} that besides being diagonalized by means of the same basis that the one for \textbf{A}, has different positive eigenvalues. This is achieved , in the case of an interaction circuit type $\boldsymbol{\Pi}$, varying their constitutive elements. Take for example a new matrix $\mathbf{\tilde{Y}}(s)$
with elements defined in terms of a resistor connected in parallel with an inductance, namely
\begin{eqnarray}\label{Y resist-induct}
\mathbf{\tilde{Y}}(s)=\frac{1}{s} \left(\begin{matrix}
\frac{1}{\mathrm{L}_a}& &0\\
0& &\frac{1}{\mathrm{L}_b}
\end{matrix}\right)+\mathbf{Y}
\end{eqnarray}
with \textbf{Y} given in \eqref{Y resist}. Here the dependence on $s$ does the job

Now the following circuit equation results
\begin{equation*}
\ddot{\mathbf{v}}(t)=-(\boldsymbol{\mho}^2 +\mathbf{DL}^{-1})\mathbf{v}(t)-\mathbf{DY\dot{v}}(t)
\end{equation*}
where one can indentify \textbf{B}=$\boldsymbol{\mho}^2 +\mathbf{DL}^{-1}$ whose eigenvalues are \{$\omega_o^2+\frac{1}{C_1\mathrm{L}_a}, \omega_o^2+\frac{1}{C_2\mathrm{L}_b}$\} while \textbf{A=DY}.
The network thus obtained is presented in Fig. \ref{RedinteraccionRPI-Lcolumnas}

\begin{figure}[!h]
\centerline{\includegraphics[totalheight=2.8cm]{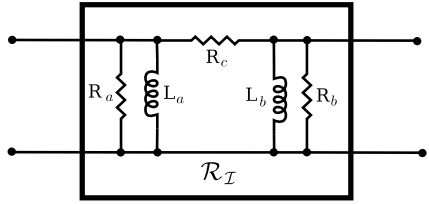}}
\caption{\footnotesize{Interaction resistive network $\mathcal{R_I}$ in topology $\Pi$ together with two columns of inductors.\label{RedinteraccionRPI-Lcolumnas}}}
\end{figure}

Now we are prepared to present the general $\mathcal{R_I}$ both connected and reciprocal that satisfies the requirement of equivalence.
The admittance matrix is given by \textbf{DY}$(s)=\boldsymbol{\alpha}+\frac{1}{s}\boldsymbol{\beta}(s)$, while the matrix elements of \textbf{Y}$(s)$ are $y_{ij}=\alpha_{ij}+\frac{\beta_{ij}}{s}$, with  $\alpha_{jj}, \beta_{jj}$ positive and with the possibility that $\alpha_{ij}, \beta_{ij}$ with $i\neq j$ can be negative.

 We the take the inverse Laplace transform to \eqref{dinamica circuital ADMITACNCIA en s} with initial condition $\dot{\mathbf{v}}(0)=\mathbf{0}$ to obtain the evolution equation
\begin{equation}\label{dinamica circuital ADMITACNCIA en t}
\ddot{\mathbf{v}}(t)+ \boldsymbol{\alpha}\dot{\mathbf{v}}(t)+(\boldsymbol{\mho}^2+\boldsymbol{\beta})\mathbf{v}(t)=0
\end{equation}
that completes the demonstration that one can get an evolution equation of the initial form \eqref{A B}. We can recognize that the matrix $\complement$ of \eqref{matriz A} is defined in terms of the following:
\begin{align*}
\mathbf{A}=\boldsymbol{\alpha},\quad \mathbf{B}=\boldsymbol{\mho}^2+\boldsymbol{\beta}
\end{align*}

Moreover, in terms of the circuit elements, one has
\begin{align}\label{Alpha}
\boldsymbol{\alpha}&=\mathbf{D}\,.\left( \begin{matrix}
\mathrm{G}_a+\mathrm{G}_c & -\mathrm{G}_c   \\
-\mathrm{G}_c    & \mathrm{G}_b+\mathrm{G}_c
\end{matrix}\right)\\
\boldsymbol{\beta}&=\mathbf{D}\,.\left( \begin{matrix}
\mathrm{L}^{-1}_a+\mathrm{L}^{-1}_c & -\mathrm{L}^{-1}_c   \\
-\mathrm{L}^{-1}_c    & \mathrm{L}^{-1}_b+\mathrm{L}^{-1}_c
\end{matrix}\right)\label{Beta}
\end{align}

The conditions $\alpha_{11}=\alpha_{22}$ and $\beta_{11}=\beta_{22}$ ensures that [$\mathbf{A,B}$]=$0$ and in this case, the characteristic polynomial is trivially factorized. Consequently, as we have shown previously, the eigenvalues of $\complement$, from \eqref{relacion de autov1} and \eqref{relacion de autov2} are
\[\lambda_j=-\tfrac{1}{2}[\mathrm{A}_j+\imath (4\mathrm{B}_j-\mathrm{A}_j^2)^\frac{1}{2}]\]
plus their complex conjugates and with $\mathrm{A}_j$ and $\mathrm{B}_j$ eigenvalues of \textbf{A} and \textbf{B} respectively. As one should have $4\mathrm{B}_j>\mathrm{A}_j^2$, one can write a first order approximation for the eigenvalues as follows
\begin{equation}\label{Identif orig}
\lambda_j=-\tfrac{1}{2}\mathrm{A}_j-\imath \sqrt{\mathrm{B}_j}
\end{equation}

By an appropriate choice of the circuit parameters that govern the eigenvalues \{$\lambda_j$\}, one can make them equal to those of the kaon matrix
$\mathbf{K}=-(\mathbf{\Gamma+\imath M})$ and build up the $S$ matrix. In this way, the isomorphism between the electric networks $\mathcal{R}$ and the kaon system $\mathcal{K}^\circ$, is established.

The general interaction reciprocal network G-L equivalent, by $\boldsymbol{\Phi}_S$, to the kaon system is sketched in Fig. \ref{Red Interacción Recíproca}.
\begin{figure}[!h]
\centerline{\includegraphics[totalheight=3cm]{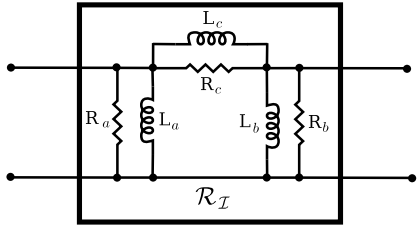}}
\caption{\footnotesize{General G-L interaction network.\label{Red Interacción Recíproca}}}
\end{figure}

As it was stated before, there exists the alternative possibility of using the currents $i_j$ instead of voltages as generalized coordinates. The corresponding dual system (type R-C), presented in Fig. \ref{Red Interacción dual}, is entirely equivalent to the kaon system as is the previous one.

\begin{figure}
\centerline{\includegraphics[totalheight=3cm]{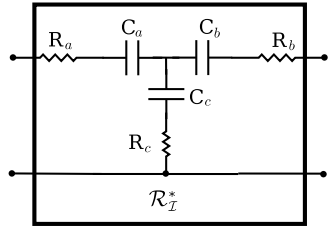}}
\caption{\footnotesize{General dual R-C interaction network.\label{Red Interacción dual}}}
\end{figure}

\subsection{Violation of CP symmetry in $\mathcal{C}$ systems, equivalent to the $\mathcal{K}^\text{o}$ system}\label{Test de violaci C -K}

The final step is to find a modified circuit in order to include the necessary information to take into account the CP violation experimentally present in the kaon system.

We have previously obtained the eigenvalues of both matrices
 $\complement$ and $\boldsymbol{\nu(K)}$. It is now necessary to obtain the $\epsilon$ parameter, or alternatively the scalar product $\boldsymbol{\mathsf{S}.\mathsf{L}}$, that measures the CP violation of the kaon system. In the basis of Pauli matrices, the effective Hamiltonian can be written as
\begin{equation}\label{mapeo completo de H}
\mathbf{H}=\left(\begin{matrix}
\frac{\mu_\text{\tiny{S}}+\mu_\text{\tiny{L}}}{2}&\frac{\mu_\text{\tiny{S}}-\mu_\text{\tiny{L}}}{2}e^{-\imath \alpha}\\
&\\
\frac{\mu_\text{\tiny{S}}-\mu_\text{\tiny{L}}}{2}e^{\imath \alpha}&\frac{\mu_\text{\tiny{S}}+\mu_\text{\tiny{L}}}{2}
\end{matrix}\right)=\mathbf{\Pi}(\mu_\text{\tiny{S}},\mu_\text{\tiny{L}},\epsilon)
\end{equation}
with $\mu_\text{\tiny{S}}=m_\text{\tiny{S}}-\imath \gamma_\text{\tiny{S}}$, $\mu_\text{\tiny{L}}=m_\text{\tiny{L}}-\imath \gamma_\text{\tiny{L}}$ being the eigenvalues of $\mathbf{H}$. Here the indices refer to $|\mathrm{K_\text{\tiny{S}}^o}\rangle$ and $|\mathrm{K_\text{\tiny{L}}^o}\rangle$, respectively. The parameter $\alpha$ is defined in terms of the $\epsilon$ parameter as
\begin{equation}
e^{\imath \alpha}=\dfrac{1-\epsilon}{1+\epsilon}
\end{equation}

Experimentally \cite{PDG 2}, the CP parameter $|\epsilon|$ is of the order of $10^{-3}$. Consequently, one can take the approximation
\begin{align*}\label{H evidencia el girador}
\mathbf{H}&=\left(\begin{matrix}
\frac{\mu_\text{\tiny{S}}+\mu_\text{\tiny{L}}}{2}&\frac{\mu_\text{\tiny{S}}-\mu_\text{\tiny{L}}}{2}\\
&\\
\frac{\mu_\text{\tiny{S}}-\mu_\text{\tiny{L}}}{2}   &\frac{\mu_\text{\tiny{S}}+\mu_\text{\tiny{L}}}{2}
\end{matrix}\right)+\left(\begin{matrix}
0&-\imath \alpha\frac{\mu_\text{\tiny{S}}-\mu_\text{\tiny{L}}}{2}\\
&\\
\imath \alpha\frac{\mu_\text{\tiny{S}}-\mu_\text{\tiny{L}}}{2}&0
\end{matrix}\right)=\mathbf{H}^{(0)}+\mathbf{H}^{(1)}
\end{align*}
Note that in this decomposition $\mathbf{H}^{(0)}$ is CP-invariant, while
\begin{equation}\label{H1}
\mathbf{H}^{(1)}=\epsilon(\Delta m-\imath\Delta \Gamma)\left(\begin{matrix}
\quad 0& & &1\\
-1& &  &0
\end{matrix}\right)
\end{equation}
 violates CP.

Remember now that the matrix $\complement$ has, in principle, 8 free parameters. Nevertheless, the condition [$\mathbf{A,B}$] $=0$ reduces this number to 6. There is a further reduction to only 5 parameters because one has that $|\Re(\epsilon)|=|\Im(\epsilon)|$ \cite{PDG 1}. We conclude that one has a sufficient number of parameters to construct an effective Hamiltonian that violates CP symmetry through the mapping between
 \textbf{H} and the matrix $\complement$.

We analyze the test of CP based on the calculation of the scalar product $\xi=\boldsymbol{\mathsf{S}.\mathsf{L}}$ written as a function of the classical circuit elements via the isomorphism $\boldsymbol{\Phi}_S$.

The information about the CP symmetry breaking is contained in the eigenvectors of \textbf{H}, i.e., in the columns of $\mathsf{Q}=(\boldsymbol{\mathsf{S}},\boldsymbol{\mathsf{L}})$.

To start with, we look for a relationship between the eigenvector basis of
 $\complement$ with that of $\boldsymbol{\nu}(\mathbf{K})$.

The real matrix $\complement$ has eigenvalues \{$\lambda_1,\lambda_1^*,\lambda_2,\lambda_2^*$\}, so its eigenvectors \{$\mathbf{w}_1,\mathbf{w}_1^*,\mathbf{w}_2,\mathbf{w}_2^*$\} form an ordered basis. They can be organized into the matrix
\[\mathbf{W}=(\mathbf{w}_1,\mathbf{w}_1^*,\mathbf{w}_2,\mathbf{w}_2^*)\]

On the other hand, the matrix $\mathsf{Q}$ contains the eigenvectors of \textbf{K};
then, the decomplexification of  $\mathsf{Q}=(\boldsymbol{\mathsf{S,L}})$, called  $\boldsymbol{\nu(\mathsf{S,L})}$ defines the transformation that takes $\boldsymbol{\nu}(\mathbf{K})$ to the real Jordan form.
To obtain  the matrix whose columns are the eigenvectors of the matrix  $\boldsymbol{\nu}(\mathbf{K})$, we find the transformation between the complex diagonal form and the real Jordan form. Note, first, that  given $ z\in\mathbb{C}$ and the unitary matrix $\Delta=\frac{1}{\sqrt{2}}\left(\begin{matrix}
\imath & \,& 1 \\
1 &\, &\imath
\end{matrix}\right)$ one has
\begin{equation*}
diag(z,z^*)=\Delta^{-1}\boldsymbol{\nu}(z)\Delta
\end{equation*}

In this way, the matrix of eigenvectors of $\boldsymbol{\nu(\mathbf{K})}$ is $\mathbf{U}=\boldsymbol{\nu}(\mathsf{Q})\boldsymbol{.}diag(\Delta,\Delta)$, that can be explicitly written as

\begin{equation*}
\mathbf{U}=\boldsymbol{\nu}(\mathsf{Q})\boldsymbol{.}diag(\Delta,\Delta)= \tfrac{1}{\sqrt{2}}\left(\begin{matrix}
\imath \mathsf{Q}_{11}& \:&\mathsf{Q}_{11}^*&\imath \mathsf{Q}_{12}& \:&\mathsf{Q}_{12}^*\\
\mathsf{Q}_{11} &\:&\imath \mathsf{Q}_{11}^*&\mathsf{Q}_{12} &\:&\imath \mathsf{Q}_{12}^*\\
\imath \mathsf{Q}_{21}& \:&\mathsf{Q}_{21}^*&\imath \mathsf{Q}_{22}& \:&\mathsf{Q}_{22}^*\\
\mathsf{Q}_{21} &\:&\imath \mathsf{Q}_{21}^*&\mathsf{Q}_{22} &\:&\imath \mathsf{Q}_{22}^*
\end{matrix}\right)=\tfrac{1}{\sqrt{2}}(\mathbf{u}_1,-\imath\mathbf{u}^*_1,\mathbf{u}_2,-\imath\mathbf{u}^*_2)
\end{equation*}

The columns of $\mathbf{U}$ are the normalized eigenvectors of $\boldsymbol{\nu(K)}$ ordered with the eigenvalues $\{\lambda_1,\lambda_1^*,\lambda_2,\lambda_2^*\}$, that clearly verify
\begin{eqnarray*}
\mathbf{u}_j\boldsymbol{.}\mathbf{u}_j=\mathbf{u}_j^\dagger \mathbf{u}_j=|\mathsf{Q}_{1j}|^2+|\mathsf{Q}_{2j}|^2=\left\lbrace \begin{matrix}
\boldsymbol{\mathsf{S}.}\boldsymbol{\mathsf{S}}=1,\: &j=1\\
\boldsymbol{\mathsf{L}.}\boldsymbol{\mathsf{L}}=1, \:&j=2
\end{matrix}\right.
\end{eqnarray*}
and $\mathbf{u}_1\boldsymbol{.}\mathbf{u}_2=
\mathsf{Q}_{11}^*\mathsf{Q}_{12}+\mathsf{Q}_{21}^*\mathsf{Q}_{22}$ implying
\begin{equation}
\mathbf{u}_1\boldsymbol{.}\mathbf{u}_2=\xi \label{xi=SL}
\end{equation}

We now relate the product $\xi=\mathbf{u}_1\boldsymbol{.}\mathbf{u}_2$ with the eigenvectors of $\complement$. As it was stated before, the classical-quantum equivalence implies that $\complement=S\boldsymbol{\nu}(\mathbf{K})S^{-1}$. Then, from
$\complement=\mathbf{W}diag(\lambda_1,\lambda_1^*,\lambda_2,\lambda_2^*)\mathbf{W}^{-1}$ and the fact that $diag(\lambda_1,\lambda_1^*,\lambda_2,\lambda_2^*)=\mathbf{U}^{-1}\boldsymbol{\nu}(\mathbf{K})\mathbf{U}$ one has $\complement=\mathbf{W}\mathbf{U}^{-1}\boldsymbol{\nu}(\mathbf{K})\mathbf{U}\mathbf{W}^{-1}$. Consequently,
\begin{equation}\label{W=SU}
\mathbf{W}=S\mathbf{U}
\end{equation}
or in other words, the eigenvectors transform with the same matrix $S$ that performs the equivalence between systems. In particular
\begin{equation}\label{w=Su}
\mathbf{w}_j=S\mathbf{u}_j
\end{equation}

From the rectangular matrices  $\mathbf{R}_{(\mathbf{w})}   =(\mathbf{w}_1,\mathbf{w}_2)$ and $\mathbf{R}_{(\mathbf{u})}=(\mathbf{u}_1,\mathbf{u}_2)$, one can define the matrix of scalar products
\begin{equation*}
\mathbf{R}^\dagger_{(\mathbf{w})}.\mathbf{R}_{(\mathbf{w})}=
\left(\begin{matrix}
\mathbf{w}_1\boldsymbol{.}\mathbf{w}_1&\quad &\mathbf{w}_1\boldsymbol{.}\mathbf{w}_2\\
\mathbf{w}_2\boldsymbol{.}\mathbf{w}_2&\quad &\mathbf{w}_2\boldsymbol{.}\mathbf{w}_2
\end{matrix}\right)=\mathbf{G}_{(\mathbf{w})}
\end{equation*}
which is the  Gramian matrix \cite{Hoffman}. In the same way one obtains
 $\mathbf{G}_{(\mathbf{u})}$. Using \eqref{w=Su} one gets
\begin{eqnarray}\label{Cambio de gramiano}
\mathbf{R}_{(\mathbf{w})}^\dagger \mathbf{R}_{(\mathbf{w})}&=\mathbf{R}_{(\mathbf{u})}^\dagger S^\dagger S \mathbf{R}_{(\mathbf{u})}
\end{eqnarray}
From $det(S)=1$ and the fact that the eigenvectors are normalized, one obtains that
 $det(\mathbf{G}_{(\mathbf{w})})=det(\mathbf{G}_{(\mathbf{u})})$ and then
$|\mathbf{w}_1\boldsymbol{.}\mathbf{w}_2|=|\mathbf{u}_1\boldsymbol{.}\mathbf{u}_2|$.
Finally, from the last equation and \eqref{xi=SL} we obtain the classical test
\begin{equation}\label{Test clasico}
|\xi|=|\mathbf{w}_1\boldsymbol{.}\mathbf{w}_2|
\end{equation}
that shows that for this analysis it is sufficient to know the eigenvectors \{$\mathbf{w}_1,\mathbf{w}_2$\} of $\complement$.

In the context of  CPT invariance, the test reduces to say that
\begin{center}
\emph{$\mathbf{w}_1\boldsymbol{.}\mathbf{w}_2\neq0$ then, $\mathrm{CP}$ is not conserved}
\end{center}

The eigenvectors of $\mathbf{W}=(\mathbf{w}_1,\mathbf{w}^*_1,\mathbf{w}_2,\mathbf{w}^*_2)$, can be written in a compact form as
\[\mathbf{w}_j=\frac{1}{\mathsf{N}_j}\boldsymbol{(} a_j^*\,\rho , a_j^*,\rho ,1\boldsymbol{)}^\intercal\]
with $a_j=\frac{\lambda_j}{\mathrm{B}_j}$, $\mathrm{B}_j$ is a $j$th eigenvalue of \textbf{B}, $\rho = \sqrt{\frac{\mathrm{C}_1}{\mathrm{C}_2}}$ and $\mathsf{N}_j=\sqrt{(\rho^2+1)(1+\frac{1}{\mathrm{B}_j})}$ chosen to ensure that $\mathbf{w}_j\boldsymbol{.}\mathbf{w}_j=1$.
Then, the classical test reduces to the computation of
\begin{equation*}
\mathbf{w}_1\boldsymbol{.}\mathbf{w}_2=\dfrac{1-\rho^2}{\mathsf{N}_1\mathsf{N}_2}\left(1+\dfrac{\lambda_1\lambda_2^*}{\mathrm{B}_1\mathrm{B}_2}\right)
\end{equation*}

We have previously stated that two families of connected and reciprocal networks equivalent to the kaon system exist. On the other hand, the existence of a dual network is guarantee by the numerical relationship
\begin{equation}\label{Rel dual CL}
\text{C}_j=\text{L}_j,\qquad \text{with } j=1,2
\end{equation}
condition that together to the requirement of CPT that $\omega_{oj}=\omega_{o}$, or equivalently that  $\text{C}_1\text{L}_1=\text{C}_2\text{L}_2$, allows one to obtain
\begin{equation*}
\text{C}_1=\text{C}_2=\text{C} \longleftrightarrow\rho=1
\end{equation*}

Consequently, one has $\mathbf{w}_1\boldsymbol{.}\mathbf{w}_2=\dfrac{1-\rho^2}{\mathsf{N}_1\mathsf{N}_2}\left(1+\dfrac{\lambda_1\lambda_2^*}{\mathrm{B}_1\mathrm{B}_2}\right)=0$, or
\begin{equation}\label{xi=0 reciproc}
|\xi|=|\mathbf{w}_1\boldsymbol{.}\mathbf{w}_2|=0,
\end{equation}
a result that ends in the conclusion:

\begin{center}
\textit{Every electrical network with the topology given in Fig. \ref{Topologia2-puertas} (linear, connected, reciprocal), and equivalent to the kaon system, represents the case in which there is invariance under CP, according to the isomorphism $\boldsymbol{\Phi}_S$.}
\end{center}

A brief review of our previous analysis shows that the only way of breaking the CP symmetry is by breaking the symmetry of the  matrices $\boldsymbol{\alpha}$ and $\boldsymbol{\beta}$. In other words, the interaction network in this case must be \textit{non-reciprocal}.

\subsection{Non reciprocal networks - Gyrators}

Remember that a network $\mathcal{R_I}$ is not reciprocal iff
$y_{ij}(s)\neq y_{ji}(s)$ for $ i\neq j$.

Due to the fact that any combination of \{R\}, \{L\}, \{C\} provides
a reciprocal network \cite{Bala}, the introduction of some new kind
of component is unavoidable. The new element called gyrator
\cite{Tellegen} does the job. This gyrator is a passive element of
two ports whose block diagram is presented in Fig.\ref{Girador}.

\begin{figure}[!h]
\centerline{\includegraphics[totalheight=3cm]{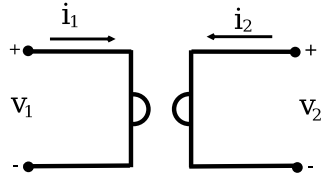}}
\caption{\footnotesize{Gyrator circuital symbol.\label{Girador}}}
\end{figure}

As a consequence of the introduction of a gyrator in a circuit, the
admittance (or impedance) matrix $\mathbb{Y}(s)$ is no longer
symmetric. In any case, for each $s$, this matrix $\mathbb{Y}(s)$
can be split as a sum of a symmetric plus an antisymmetric matrix
\cite{Hoffman}. In terms of electric networks we write
$\mathbb{Y}(s)=\mathbf{Y}(s)+\mathbf{Y}_{\thicksim}(s)$ ,
i.e., the sum of a reciprocal matrix plus a non reciprocal one. Here
$\mathbf{Y}(s)=\mathrm{C}[\boldsymbol{\alpha}
+\boldsymbol{\beta}(s)]$, with $\boldsymbol{\alpha}$ and
$\boldsymbol{\beta}$ symmetric. \cite{Carlin Giordano2}

Taking into account that the CP
violation is measured by a parameter of the order of
$|\epsilon|<10^{-3}$, one should guarantee that the matrix
$\mathbf{Y}_{\thicksim}(s)$ is small. Consequently,
$\mathbf{Y}_{\thicksim}(s)$ acts as a small perturbation.
Moreover, $\mathbf{Y}_{\thicksim}$ is real due to \eqref{dinamica
circuital ADMITACNCIA en s} and then has the form
\begin{equation}\label{Matriz adm del girador}
\mathbf{Y}_{\thicksim}:=\text{C}\boldsymbol{\alpha}_g=\left(\begin{matrix}
\quad 0 &\quad &g\\
-g& \quad &0
\end{matrix}\right)
\end{equation}

This is precisely the admittance representation of the gyrator
\cite{Tellegen}, where $g$ ($>0$) is its conductance that we call
$\mathbf{Y}_{\thicksim}=\mathbf{Y}_g$, and $ \boldsymbol{\alpha}_g $ is the corresponding antisymmetric term of new matrix $\boldsymbol{\alpha+\alpha}_g$

From this, one can write the new classical non-reciprocal matrix $\mathbb{C}$

\begin{align}\label{Matriz C con girador}
\mathbb{C}=\left(\begin{matrix}
\quad \mathbf{0} &\quad &\mathbf{1}\\
-\boldsymbol{\boldsymbol{\mho}^2-\beta}& \quad &-\boldsymbol{\alpha}-\boldsymbol{\alpha}_g
\end{matrix}\right)&=\left(\begin{matrix}
\quad \mathbf{0} &\quad &\mathbf{1}\\
-\boldsymbol{\boldsymbol{\mho}^2-\beta}& \quad &-\boldsymbol{\alpha}
\end{matrix}\right)+\left(\begin{matrix}
\mathbf{0} &\quad &\mathbf{0}\\
\mathbf{0}& \quad &-\boldsymbol{\alpha}_g
\end{matrix}\right)\nonumber
\end{align}
then is  defined
\begin{equation}
\mathbb{C}=\complement + \complement_g\label{Nueva matriz clasica}
\end{equation}
where $\boldsymbol{\beta}$ and $\boldsymbol{\alpha}$ are defined in \eqref{Alpha}.

Notice that the non-reciprocal matrix $\boldsymbol{\alpha}_g$ is proportional to the particular term of the effective Hamiltonian \eqref{H1} of the quantum system that violates CP
\begin{equation*}
\boldsymbol{\alpha}_g\propto\mathbf{H}^{(1)}
\end{equation*}

From \eqref{Nueva matriz clasica} one can say that the gyrator works
as a perturbation of the original CP invariant network, similarly to
the term $\mathbf{H}^{(1)}$ in the quantum system. For the new
classical matrix $\mathbb{C}$ \eqref{Nueva matriz
clasica}, the characteristic polynomial takes the form
\begin{align}
p_{_{\mathbb{C}}}(z)&=p_{_{\complement}}(z)+\left(\frac{g}{\scriptsize{\text{C}}}\right)^2z^2\label{Pol C no rec}
\end{align}
that shows that the perturbative first order is quadratic in $g$. Clearly, one has
to expect that this parameter $g$ should be of the order of $\epsilon$. In the perturbation theory scheme, we consider the expansion of the roots  \{$\eta$\} of $p_{_{\complement{\sim r}}}(z)$, in terms of those roots \{$\lambda$\} of $p_{_{\complement}}(z)$

\begin{equation}\label{lambda pert}
\eta(g)=\lambda+\sum_{k\in\mathbb{N}} \lambda_kg^k
\end{equation}

To compute the relevant parameter $\xi$ we consider the eigenvectors
corresponding to the two eigenvalues \{$\lambda_1,\lambda_2$\} of
the third quadrant. Again one writes for the eigenvectors
\{$\boldsymbol{\varpi}$\} of the matrix $\mathbb{C}$ as an expansion in
terms of those \{\textbf{w}\} of $\complement$
\begin{equation}\label{w pert}
\boldsymbol{\varpi}(g)=\mathbf{w}+\sum_{k\in\mathbb{N}} \mathbf{w}_{\sim}^{(k)}g^k
\end{equation}
that verify
\begin{equation}\label{Complement no reci}
\mathbb{C}\boldsymbol{\varpi}(g)=\eta(g)\boldsymbol{\varpi}(g)
\end{equation}

 To first order in $g$ one has
\begin{equation}\label{w pert 1}
\boldsymbol{\varpi}(g)=\mathbf{w}+g\mathbf{w}_{\sim}
\end{equation}
that ends in
\begin{equation}
g(\lambda\mathbf{1}-\complement)\mathbf{w}_\sim = \complement_g \mathbf{w}
\end{equation}

Computing $|\xi|=|\boldsymbol{\varpi}_1(g)\boldsymbol{.}\boldsymbol{\varpi}_2(g)|=|\mathbf{w}_1\centerdot\mathbf{w}_2+g(\mathbf{w}_{1\sim}\centerdot\mathbf{w}_{2}
+\mathbf{w}_1\boldsymbol{.}\mathbf{w}_{2\sim})+g^2\mathbf{w}_{1\sim}\centerdot
\mathbf{w}_{2\sim}|$ and considering that $\mathbf{w}_1\centerdot\mathbf{w}_2=0$, the results
is
\begin{equation}
|\xi|=g|\mathbf{w}_{1\sim}\centerdot\mathbf{w}_{2}
+\mathbf{w}_1\boldsymbol{.}\mathbf{w}_{2\sim}|
\end{equation}
showing that, as announced, the breaking of CP symmetry is measured by the gyrator conductance $g$.

The important point to remark is that there exists a topological structure that physically realizes this circuit. The family of networks of interaction that violates CP, being equivalent to the kaon system and invariant under CPT is
\[\mathcal{R_I}=\mathcal{R}_{r} ||\:\mathsf{G}\]
where $\mathcal{R}_{r}$ is Fig. \ref{Red Interacción Recíproca}; $\mathsf{G}$ represents a gyrator and both are connected in parallel. The network doing the job is shown in Fig. \ref{Familia de Redes que violan CP}
\begin{figure}[!h]
\centerline{\includegraphics[totalheight=3cm]{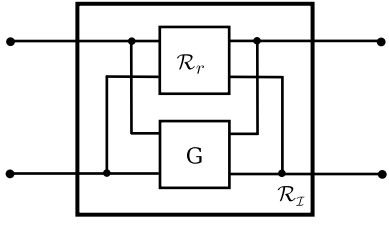}}
\caption{\footnotesize{Interaction network family type \textbf{Y}, equivalent to the kaon system including CP violation.\label{Familia de Redes que violan CP}}}
\end{figure}

Clearly, if one consider the port currents instead of the voltages as generalized coordinates, the corresponding family of interaction networks includes the dual circuits, namely
 \[\mathcal{R^*_I}=\mathcal{R}^*_{r}\oplus\mathsf{G^*}\]
where $\mathcal{R}^*_{r}$ verifies CP symmetry (Fig. \ref{Red Interacción dual}), $\mathsf{G^*}$ is the dual of a gyrator (also a gyrator \cite{Bala}) and they are connected in series. This is sketched in Fig. \ref{Familia de Redes que violan CP dual}
\begin{figure}[!h]
\centerline{\includegraphics[totalheight=3cm]{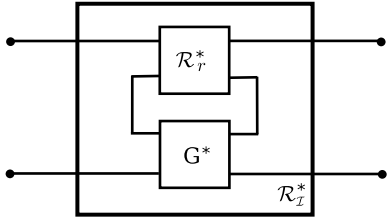}}
\caption{\footnotesize{Interaction dual network family type \textbf{Z}, equivalent to the kaon system including CP violation.\label{Familia de Redes que violan CP dual}}}
\end{figure}

To complete the mapping of \textbf{H} into the matrix elements of $\complement$, we should specify explicitly its eigenvalues and the value of $\epsilon$. Notice that we have imposed that $\alpha_{11}=\alpha_{22}$ and $\beta_{11}=\beta_{22}$, in \eqref{Alpha} and \eqref{Beta}. This implies that G$_a$=G$_b$ and L$_a$=L$_b$. Moreover, the existence of dual networks implies that $\boldsymbol{\alpha}$ and $\boldsymbol{\beta}$ have to be symmetrical. Then
\begin{align}
\boldsymbol{\alpha}=\left( \begin{matrix}
\tau_a^{-1}+\tau_c^{-1} & -\tau_c^{-1} \\
-\tau_c^{-1}    &\tau_a^{-1}+\tau_c^{-1}
\end{matrix}\right),\quad
\boldsymbol{\beta}=\left( \begin{matrix}
\omega_a^2+\omega_c^2 & -\omega_c^2\\
-\omega_c^2 &\omega_a^2+\omega_c^2
\end{matrix}\right)\label{Beta final}
\end{align}
where $\tau_a=\frac{\mathrm{C}}{\mathrm{G}_a}$, $\tau_c=\frac{\mathrm{C}}{\mathrm{\mathrm{G}_a}}$, $\omega_a^2=\frac{1}{\mathrm{CL}_a}$ and $\omega_c^2=\frac{1}{\mathrm{CL}_c}$.
The eigenvalues of \textbf{A} and \textbf{B} have the form $\mathrm{A}_{11}\pm\mathrm{A}_{12}$ and $\mathrm{B}_{11}\pm\mathrm{B}_{12}$ respectively. Finally, the eigenvalues of \textbf{K}, $-\gamma_j-\imath m_j$, are equal to the two eigenvalues of $\complement$ belonging to the third quadrant. The connection with the circuit elements coming from \eqref{Identif orig} is
\begin{align}
\gamma_\text{\tiny{L}}=\tau^{-1}_a,&\quad
\gamma_\text{\tiny{S}}=\tau^{-1}_a+2\tau^{-1}_c \label{Identificación1}\\
m_\text{\tiny{L}}=\sqrt{\omega_o^2+\omega_a^2}, & \quad
m_\text{\tiny{S}}=\sqrt{\omega_o^2+\omega_a^2+2\omega_c^2} \label{Identificación2}
\end{align}

On can also write the difference of masses and widths
\begin{align}\label{Dif masas y gamas}
\Delta m&=m_{\text{\tiny{S}}}-m_{\text{\tiny{L}}}\simeq \sqrt{2}\omega_c\\
\Delta \Gamma &=\Gamma_{\text{\tiny{S}}}-\Gamma_{\text{\tiny{L}}}=2\tau_c^{-1}
\end{align}

As $\Delta m \sim \Delta\Gamma$ and $\Delta\Gamma\simeq\gamma_{\text{\tiny{S}}}\gg \gamma_{\text{\tiny{L}}}$ \cite{PDG 1} it must be that
\begin{equation}
\omega_c \tau_c=\sqrt{2}
\end{equation}

Under the hypothesis that the small perturbation on the initial reciprocal circuit does not strongly alter the proper frequencies of the free system, one can get easily the  $\epsilon$ parameter. By comparing with the corresponding Laplace-transformed differential equations of both dynamical systems \eqref{Sch} and \eqref{dinamica circuital ADMITACNCIA en s}, considering the initial condition $\mathbf{i}_\mathrm{L}(0)=\mathbf{0}$ in \eqref{Corriente inductor} for the classical equivalent system. Then, from \eqref{Sch} and \eqref{dinamica circuital ADMITACNCIA en s} we obtain
\begin{align}
s\Psi(s)-\psi(0)&=-\imath\mathbf{H}\Psi(s)\\
s\mathbf{V}(s)-\mathbf{v}(0)&=\mathcal{M}(s)\mathbf{V}(s)
\end{align}
by definition $\Psi(s):=\mathcal{L}\{\psi(t)\}$ and $\mathcal{M}(s):=-\left(\frac{1}{s}\mho^2+\mathbf{DY}(s)\right)$. The above hypothesis of the small perturbation implies $\mathcal{M}(s)\simeq\mathcal{M}(-\imath\omega_o)$, we obtain the equivalence
\begin{equation*}
\mathbf{H}\sim\imath\mathcal{M}_{(-\imath\,\omega_o)}=\mho-\imath\mathbf{DY}_{(-\imath\,\omega_o)}
\end{equation*}

From  $\frac{1}{\text{C}}\mathbf{Y}(s)=\boldsymbol{\alpha} + \frac{1}{s}\boldsymbol{\beta}$ with \textbf{A}=$\boldsymbol{\alpha}$ and \textbf{B}=$\boldsymbol{\mho^2+\beta}$ obtain
\begin{equation*}
\mathbf{H}\boldsymbol{\sim}\scriptstyle{\frac{\scriptstyle 1}{\scriptstyle \omega_o}}\displaystyle \mathbf{B}-\imath \mathbf{A}=\left(
\begin{matrix}
\omega_o-\imath\: \frac{y_{11}(-\imath\omega_o)}{\mathrm{C}}&-\imath\: \frac{y_{12}(-\imath\omega_o)}{\mathrm{C}}\\
\\
-\imath\: \frac{y_{21}(-\imath\omega_o)}{\mathrm{C}}&\omega_o-\imath\: \frac{y_{22}(-\imath\omega_o)}{\mathrm{C}}\\
\end{matrix}\right)
\end{equation*}

On the other hand in the case of neutral kaons system, from \eqref{Epsilon posta} we obtain \cite{Lee}

\begin{equation}\label{Epsilon}
\epsilon \simeq\left. \frac{\mathrm{H}_{12}-\mathrm{H}_{21}}{\sqrt{\mathrm{H}_{12}\mathrm{H}_{21}}}\right.
\end{equation}

Therefore the equivalent network epsilon parameter is given for

\begin{eqnarray}\label{Epsilon circuito2}
\epsilon  \simeq\frac{y_{21}(-\imath \omega_o)-y_{12}(-\imath \omega_o)}{\sqrt{y_{12}(-\imath \omega_o)y_{21}(-\imath \omega_o)}}
\end{eqnarray}
that certainly vanishes when $\mathcal{R_I}$ is reciprocal.

Taking into account the non-reciprocal case \textbf{Y}$(s)$=\textbf{Y}$_{r}(s)$+\textbf{Y}$_{\sim r}(s)$ whose elements are $y_{ij}(s)= y_{ij}^{(r)}(s)+(-)^j g\delta_{ij}$ and that 	$y^{(r)}_{12}(-\imath \omega_o)=-\mathrm{G}_c- \dfrac{\imath}{\omega_o\mathrm{L}_c}$ one obtains
\begin{eqnarray*}
\epsilon  \simeq-\frac{2g}{\sqrt{(\mathrm{G}_c+ \frac{\imath}{\omega_o\mathrm{L}_c})^2-g^2}}
\end{eqnarray*}
that to order $g$ reduces to
\begin{eqnarray}\label{epsilon apros}
\epsilon  \simeq-\frac{2g}{\mathrm{G}_c+ \dfrac{\imath}{\omega_o\mathrm{L}_c}}
\end{eqnarray}
showing again clearly that the gyrator conductance $g$ governs the CP violation in the classical $^\backprime$equivalent dynamic$’$.
\vspace*{1cm}
\section{Conclusions}\label{Conclusiones}

We have been able to extend the character of the nice observation by
Rosner \cite{Rosner 1} about the characteristic matrix of a given
circuit having a similar aspect to the effective Hamiltonian of the
kaon system. We have gone beyond an \textit{analogy}, to obtain an
\textit{equivalence}, stricto sensu, between the
$\mathrm{Schr\ddot{o}dinger}$ dynamics of a quantum system with a
finite number of basis states  and the classical dynamics of
electric networks. The equivalence we present is an isomorphism that
connects in univocal way both dynamical systems.

Our conclusions are based upon the concept of $^\backprime$equivalent dynamic$’$ defined starting from general mathematical
considerations. Since the $\mathrm{Schr\ddot{o}dinger}$ dynamics is
defined on a complex space, while the classical dynamics of interest
implies real evolution equations, we have used the
decomplexification procedure that allows us to work in a common
space. In this way we define the general classes of classical
$\mathcal{C}$ and quantum $\mathcal{Q}$ systems where the
equivalence is valid and simultaneously, we build up the isomorphism
$\Phi_S$, that takes $\mathcal{C}$ into $\mathcal{Q}$ and includes
the general correspondence between states of both systems.

We have presented here the particular case of neutral kaons, as the
quantum systems, particularly interested in the aspects of
$\mathrm{CP}$ invariance in the context of validity of
$\mathrm{CPT}$ symmetry.

The concept of circuital duality allows us to obtain two
equivalent electrical representations of the same classical
differential equation. This makes easy the choice of the
parameters that govern the $\mathrm{CP}$ or $\mathrm{T}$ violation
in the network.

The class of electric networks $\mathcal{R}$ is univocally related
to the kaon system $\mathcal{K}^\mathrm{o}$ since we have found the
complete map between the matrix elements of the effective
Hamiltonian of kaons and those elements of the classical dynamics of
the networks. In fact, there exists a one to one relationship
between the states $|\mathrm{K^o}\rangle$ and
$|\bar{\mathrm{K}}^\mathrm{o}\rangle$ and port voltages, or
currents, of the electric network.

We have presented a formal classical test of the $\mathrm{CP}$
invariance that is a reflection of the quantum test. From this test,
together with the concept of dual network, one concludes that any
violation of the  $\mathrm{CP}$ (or  $\mathrm{T}$) symmetry is
directly related to the presence of non reciprocity in the network.
The observable associated to the violation of $\mathrm{T}$
invariance at quantum level is related, in our realization, to
the conductance of a gyrator. The gyrator is a two-port, non-reciprocal, passive network without losses, and violates the classical symmetry $\mathrm{T}$. We then end up with a network completely equivalent to the kaon system, that allows one to represent the
relevant parameters of the quantum system in terms of circuit
components. The interaction between both L$-$C subnetworks gives
rise to a shift in the proper initial free frequencies, in the same
way as the masses of kaons. Moreover, the presence of proper
relaxation times of the circuit are associated to the mean lives of
K-\textit{short} and K-\textit{long}.

It is possible to generalize these ideas to  other quantum systems immediately, to  map the quantum Hamiltonian  and rewrite it in term of the elements of equivalent  classical system, along with the study of the underlying symmetries.

\section*{Acknowledgments}

We warmly thank Professor Jonathan Rosner for his encouraging comments and a careful reading of the manuscript that has improved our presentation in every aspect.
This research was supported by CONICET and ANPCyT, Argentina.


\begin{thebibliography}{99}

\bibitem{Rosner 1}{J.L.Rosner, \emph{Tabletop time-reversal violation}, Am.J.Phys.\textbf{64} (8), 982-985, (1996)}

\bibitem{Kostelecky-Roberts}{V. Alan Kosteleck$\acute{\mathrm{y}}$, Ágnes Roberts, \emph{Analogue models for T and CPT violation un neutral-meson\\ oscillations}, Phys. Review D \textbf{63}, 096002-1 (2001)}

\bibitem{Cocolicchio 1}{D.Cocolicchio, \emph{The Classical Analogue of CP-Violation},   Foundations of Physics Letters \textbf{11} (1), 23-39, (1998)}

\bibitem{Feynman 1}{R.P. Feynman, \emph{The Feynman Lectures on Physics, Volume 3 Quantum Mechanics}, Pearson | Addison-Wesley, (1964)}

\bibitem{Hirsch}{M.W. Hirsch, S. Smale, \emph{Differential Equations, Dynamical System and Linear Algebra}, Academic Press Inc, (1974)}

\bibitem{Arnold1}{V. I. Arnold, \emph{Ordinary Differential Equations}, The MIT Press, (1995), 9$^\mathrm{th}$ edition}

\bibitem{Lee}{T.D. Lee, \emph{Particle Physics and Introduction to Field Theory}, Harwood Academic Publishers, (1981)}

\bibitem{PDG 1}{M. Antonelli, G. D’Ambrosio (Particle Data Group), \emph{CPT Invariance tests in neutral Kaon decay}, Updated October (2009)}

\bibitem{PDG 2}{L. Wolfenstein, T.G. Trippe, C.-J. Lin (Particle Data Group), \emph{Tests of conservation laws}, Updated June (2008)}

\bibitem{PDG 3}{C. Amsler et al. (Particle Data Group), PL \textbf{B667}, 1, (2008)}

\bibitem{Bala}{N. Balabanian, T.A. Bickart, S. Seshu, \emph{Electrical network theory}, Wiley, (1969)}

\bibitem{Carlin Giordano1}{H. Carlin, A. Giordano, \emph{Network Theory: An Introduction to Reciprocal and Nonreciprocal Circuits}, Prentice Hall, (1964)}

\bibitem{Tellegen}{B.D.H. Tellegen, \emph{The Gyrator, a new electric network element}, Philips Res. Rept. \textbf{3}, 81-101, (1948)}

\bibitem{Marsden1}{Marsden, Ratiu, Abraham, \emph{Manifolds, Tensor Analysis, and Applications}, Springer T (2003)}

\bibitem{Marsden2}{Marsden, Ratiu, \emph{Introduction to Mechanics and Symmetry}, Springer-Verlag (1994)}

\bibitem{Dirac}{P.A.M.Dirac \emph{The principles of Quantum Mechanics}, Oxford University Press, (1958), 4$^{\circ}$ edición}

\bibitem{Galindo}{A.Galindo, P. Pascual \emph{Quantum Mechanics}, Springer-Verlag, (1990)}

\bibitem{Silvester}{J. R. Silvester \emph{Determinants of Block Matrices}, Maths. Gazette \textbf{84}, 460-467, (2000), \textit{http://www.mth.kcl.ac.uk/$\sim$jrs/gazette/blocks.pdf}}

\bibitem{Caruso}{M. G. Caruso \emph{Isomorphism between the Schr$\ddot{\text{o}}$dinger Dynamics and Classical Dynamics}, Work of degree thesis}

\bibitem{Wigner-Weisskopf}{V.F. Weisskopf, E.P. Wigner, Z. Physik. \textbf{63} (54), 1930,\textbf{65} (18), (1930)}

\bibitem{Froggatt Nielsen}{C. D. Froggatt,H. B. Nielsen, \emph{Origin of Symmetries}, World Scientific, (1991)}

\bibitem{Streater Wightman}{R.F. Streater, A.S. Wightman, \emph{PCT, Spin and Statistics, and all that}, Princeton University Press, Third Edition (1980)}

\bibitem{Luders}{G. L$\mathrm{\ddot{u}}$ders, \emph{Proof of TCP Theorem}, Ann. Phys. \textbf{2},1 (1957)}

\bibitem{Sachs}{R.G. Sachs, \emph{The Physical of Time Reversal}, The University of Chicago Press | Chicago and London, (1987)}

\bibitem{Lee Oehme Loy}{T.D. Lee, R. Oehme, C.N. Yang, \emph{Remarks on Possible Noninvariance under Time Reversal and Charge Conjugation}, Phys.Rev. \textbf{106} (2) 340-345, (1957)}

\bibitem{Christenson}{ J.H. Christenson, J.W. Cronin, V.L. Fitch, R. Turlay \emph{Evidence for the $2\pi$ Decay of the K$_2^{\circ}$ Meson}, Phys. Rev. Lett. \textbf{13}, 138, (1964)}

\bibitem{Sakurai}{J.J. Sakurai, \emph{Invariance Principles and Elementary Particles}, Princeton University Press, (1964)}

\bibitem{Whitney}{H.Whitney, \emph{Nonseparable and Planar Graphs}, Trans. Amer. Math. Soc.,\textbf{34} (1932), 339–362.}

\bibitem{Hoffman}{K. Hoffman, R. Kunze, \emph{Linear algebra}, Prentice Hall Internacional, (1973)}

\bibitem{Carlin Giordano2}{H. Carlin, \emph{On the Physical Realizability of Linear Non-Reciprocal Networks}, Prentice Hall, (1964)}

\end{thebibliography}
\end{document}